\documentclass{original-ptephy}

\makeatletter
\@addtoreset{equation}{section}

\makeatother

\newcommand{\be}{\begin{equation}}
\newcommand{\ee}{\end{equation}}
\newcommand{\bea}{\begin{eqnarray}}
\newcommand{\eea}{\end{eqnarray}}
\newcommand{\beann}{\begin{eqnarray*}}
\newcommand{\eeann}{\end{eqnarray*}}
\newcommand{\nn}{\nonumber}
\newcommand{\ba}{\begin{array}}
\newcommand{\ea}{\end{array}}

\DeclareMathOperator{\Tr}{Tr}
\DeclareMathOperator{\Det}{Det}
\DeclareMathOperator{\diag}{diag}
\DeclareMathOperator{\ind}{ind}

\newcommand{\e}{\epsilon}

\newcommand{\C}{\mathbb{C}}

\newcommand{\del}{\partial}
\newcommand{\D}{{\mathcal D}}

\newcommand{\B}{{\mathcal B}}
\newcommand{\F}{{\mathcal F}}

\newcommand{\Op}{{\mathcal O}}

\newcommand{\Kahler}{K\"ahler }

\newcommand{\zb}{{\bar{z}}}

\newcommand{\Volume}{{\rm Vol}}
\newcommand{\A}{{\hat{\mathcal A}}}

\begin{document}

\title{
Localization Method for Volume of Domain-Wall Moduli Spaces
}

\author{
\name{\fname{Kazutoshi} \surname{Ohta}}{1},
\name{\fname{Norisuke} \surname{Sakai}}{2}, and
\name{\fname{Yutaka} \surname{Yoshida}}{3}}

\address{\affil{1}{Institute of Physics, Meiji Gakuin University, Yokohama 244-8539, Japan}
\affil{2}{Department of Mathematics, Tokyo Woman's Christian University, Tokyo 167-8585, Japan}
\affil{3}{High Energy Accelerator Research Organization (KEK), Tsukuba, Ibaraki 305-0801, Japan}
\email{${}^1$kohta@law.meijigakuin.ac.jp, ${}^2$norisuke.sakai@gmail.com, ${}^3$yyoshida@post.kek.jp}
}

\begin{abstract}%
Volume of moduli space of non-Abelian BPS domain-walls
is exactly obtained in $U(N_c)$ gauge theory with $N_f$ matters.
The volume of the moduli space is formulated, without an explicit metric,
by a path integral
under constraints on BPS equations. 
The path integral over fields reduces to a finite dimensional contour integral
by a localization mechanism.
Our volume formula satisfies a Seiberg like duality
between moduli spaces of the $U(N_c)$ and $U(N_f-N_c)$ non-Abelian BPS domain-walls
in a strong coupling region.
We also find a T-duality between domain-walls and vortices on a cylinder.
The moduli space volume of non-Abelian local ($N_c=N_f$) vortices on the cylinder
agrees exactly with that on a sphere. 
The volume formula
reveals
various geometrical properties of the moduli space.
\end{abstract}


\maketitle

\section{Introduction}

A moduli space of Bogomol'nyi-Prasad-Sommerfield (BPS) 
solitons, which is a space of parameters 
describing positions, orientations and sizes, is 
important to understand properties of BPS solitons 
themselves.
For example, metric of the moduli space is important to 
see scatterings among the BPS solitons.

Volume of the moduli space is essentially obtained from 
an integral of a volume form, which is constructed by 
the metric, over the moduli space. 
A local structure of the moduli space is smeared out by 
the volume integration, but the volume of the moduli 
space still has significant informations on dynamics of 
the BPS solitons. 
The volume of the moduli space is directly proportional 
to a thermodynamical partition function 
of many body system of the BPS solitons.
Thermodynamics of vortices is investigated by evaluation 
of the volume of the moduli space 
\cite{Manton:1993tt,Shah:1993us,Manton:1998kq,Nasir:1998kt,Manton:2004tk}.

The volume of the moduli space of the BPS solitons also 
tells us non-perturbative dynamics in 
supersymmetric gauge field theories.
Nekrasov has shown that one of non-perturbative 
corrections in ${\mathcal N}=2$ supersymmetric gauge 
theories in four dimensions can be obtained from a 
volume of moduli space of self-dual 
Yang-Mills instantons \cite{Nekrasov:2002qd}
by using a {\it localization method}, developed in 
\cite{Moore:1997dj,Gerasimov:2006zt}.
The localization method recently becomes  more important 
to investigate the non-perturbative dynamics 
of supersymmetric gauge theories through exact partition 
functions. 
The exact partition function of supersymmetric gauge 
theory is essentially proportional to the volume of the 
moduli space of the BPS solitons, which produce the 
non-perturbative corrections.

It is very difficult to construct an explicit metric of 
the moduli space of the BPS solitons in 
general\cite{Manton:2004tk}, 
so the calculation of the volume of the moduli space is 
difficult, too. 
However, we do not need an explicit metric on the moduli 
space to evaluate the volume in the localization method.
This fact comes from integrability and supersymmetry 
behind the BPS solitons. 
Indeed, the supersymmetry is closely related to {\it 
equivariant cohomology}, which plays an important role 
in mathematical formulation of the localization method. 
Then, the localization method is very useful to calculate 
the volume of the moduli space and extends a range of 
applicable cases in the volume calculation of the BPS 
soliton moduli space.

The advantage of the localization method in 
calculating the volume has been shown in the 
calculation of the volume of the instanton moduli 
space, which gives the non-perturbative corrections 
in four-dimensional supersymmetric gauge 
theory \cite{Nekrasov:2002qd}. 
And then, the localization method is applied to evaluate 
the volume of the moduli space of the non-Abelian BPS 
vortices \cite{Miyake:2011yr}. 
The results from the localization method perfectly agree 
with the previous results using the other method, and 
we could extend to more complicated systems, 
where the metric of the moduli space is not explicitly 
known.

In this paper, we calculate the volume of the moduli 
space of the non-Abelian BPS domain-walls, 
which is described by first order differential equations 
for matrix- and vector-valued variables, where the 
matrices are in adjoint representations of $U(N_c)$ and 
$N_f$ sets of the vectors are in fundamental 
representations of $U(N_c)$. 
We consider the BPS equations of the domain-walls
on a finite line interval with boundaries.
Solutions of the BPS domain-wall equations depend on 
boundary conditions. 
So we need to carefully treat the boundary conditions 
to consider the moduli spaces of the BPS domain-walls.
The differential equations of the domain-walls can be 
regarded as a BPS equations in supersymmetric gauge 
theory with $U(N_c)$ gauge group and $N_f$ flavors 
(matters) in the fundamental representation.
The domain-walls are soliton like object with 
co-dimension one in supersymmetric gauge theory.
We are interested in the moduli space of the BPS 
equations only, so we do not assume an explicit 
supersymmetric system in the calculation of the volume.

We utilize the localization method associated with the 
equivariant cohomology in mathematics 
in order to evaluate the volume of the moduli space of 
the BPS domain-walls.
The localization method is essentially equivalent to an 
evaluation of a field theoretical partition function 
of some constrained system. A path integral of the 
partition function is restricted on the moduli space 
of the domain-walls.
We again emphasize that we need the constraints of the 
BPS equations, but do not need an explicit metric of 
the moduli space in this localization method.

The path integral which gives the volume of the moduli 
is localized at fixed points of a symmetry, 
which is a part of the supersymmetry. This symmetry is 
called a Becchi-Rouet-Stora-Tyutin (BRST) symmetry and 
related to the equivariant cohomology.
In the evaluation of the path integral, it is necessary 
to know the number of zero modes of the fields. 
We find that the number of the zero modes is determined 
by the boundary conditions, and is given by a
Callias like index theorem with boundary. 
After counting the zero modes explicitly, we find that 
the path integral reduces to a usual contour integral 
and a simple formula is obtained for the volume of the 
moduli space of the BPS domain-walls. 
 For non-Abelian gauge theories, 
we find that the contour integral reduces to 
a sum of products of the Abelian gauge theories with 
non-trivial signs. 
The sign of each product in the sum could not be 
determined by the localization method itself. 
We assume that the signs is determined by a 
topological index (intersection number) of the profile 
of the solution. 
Then, the sum of products is expressed by a determinant 
of a simple matrix depending on the boundary conditions.

In order to check our volume formula for the moduli 
space of the BPS domain-walls, we discuss dualities 
between various systems of the domain-walls. 
First of all, we investigate the duality between the 
moduli spaces of the non-Abelian BPS 
domain-walls in the strong coupling (asymptotic) region.
We find that the moduli spaces of the domain-walls of 
$U(N_c)$ and $U(\tilde{N}_c)$ differ from each other 
in general, but if $\tilde{N}_c$ is given by $N_f-N_c$, 
then we expect that the moduli spaces (and its volume) 
coincide with each other in the strong coupling region 
\cite{Isozumi:2004va,Antoniadis:1996ra}.
We can conclude that our results agrees with the expected 
dualities.
Secondly, we show that there exists a T-dual relation 
between the domain-walls and vortices on a cylinder 
\cite{Eto:2004vy}.
The domain-walls and vortices have different 
co-dimensions, but if we consider the domain-walls 
winding along a circle direction of the cylinder, 
the volume of the moduli space can be regarded as 
that of the moduli space of the vortices 
on the cylinder \cite{Eto:2007aw}.
The winding number of the domain-walls corresponds to a 
vortex charge. 
We find that the volume of the moduli space of the 
vortices on the cylinder coincides with that of the 
vortices on the sphere if $N_c=N_f$ 
(non-Abelian local vortex). 
These non-trivial duality relations support that our 
volume formula for the moduli space of the BPS 
domain-walls correctly works.

This paper is organized as follows:
In the next section, we explain a general argument on the 
volume calculation of the moduli space of the BPS 
equations. 
We introduce a 
path integral over the constrained system 
to evaluate the volume without the explicit metric.
In section 3, we evaluate the path integral to see 
that it is localized at fixed points of the BRST 
symmetry, and reduces to a simple contour integral. 
In section 4, we explicitly evaluate the contour 
integral for various examples of domain-walls in 
Abelian and non-Abelian gauge theories. 
In order to check our results for the volume of the 
moduli space of the BPS domain-walls, 
we consider two kinds of dualities of the moduli spaces 
in section 5 and 6.
The last section is devoted to conclusion and discussion.

\section{Volume of Moduli Space}

We take the $U(N_c)$ gauge theory with the 
gauge field $A_\mu$, together 
with a real scalar field $\Sigma$ in the adjoint 
representation and $N_f$ complex scalar field $H_r^A, 
r=1, \cdots N_c, A=1, \cdots, N_f$ 
in the fundamental representation. 
The gauge coupling and the Fayet-Iliopoulos (FI) 
parameter are denoted as $g$ and $c$, respectively. 
Let us consider the BPS equations for domain-walls 
\cite{Isozumi:2004jc,Isozumi:2004va,Eto:2006ng} 
in a finite interval $y\in [-\frac{L}{2},\frac{L}{2}]$:
\bea
\mu_r &\equiv&\D_y \Sigma -\frac{g^2}{2}\left(
c {\bf 1}_{N_c} - HH^\dag\right)=0,
\label{BPS1}\\
\mu_c&\equiv&\D_y H + \Sigma H - H M=0,
\label{BPS2}\\
\mu_c^\dag&\equiv&\D_y H^\dag + H^\dag \Sigma - M H^\dag=0,
\label{BPS3}
\eea
where $\Sigma$, $H$ and $H^\dag$ are $N_c\times N_c$, 
$N_c\times N_f$ and $N_f\times N_c$ matrix-valued 
functions of $y$, respectively, 
and the covariant derivatives are defined by 
$\D_y \Sigma = \del_y\Sigma + i[A_y,\Sigma]$,
$\D_y H = \del_yH+iA_y H$ and 
$\D_y H^\dag = \del_y H^\dag -i H^\dag A_y$. 
The mass matrix $M$ is taken to be diagonal as 
$M=\diag(m_1,m_2,\ldots,m_{N_f})$ and ordered as 
$m_1<m_2<\cdots<m_{N_f}$ without loss of generality. 

Domain-wall solutions are defined by specifying vacuum 
at the left and right boundaries. 
Vacua of the system are labeled by choosing $N_c$ out 
of $N_f$ flavors, 
\cite{Isozumi:2004jc,Isozumi:2004va,Eto:2006ng} 
such as $(A_1, \cdots, A_{N_c})$, 
with $A_1< A_2 < \cdots A_{N_c}$. 
Let us consider domain-wall solutions connecting 
the vacuum $(A_1, \cdots, A_{N_c})$ at the left boundary 
$y=-L/2$ and the vacuum $(B_1, \cdots, B_{N_c})$ 
at the right boundary $y=L/2$. 
For finite intervals, we demand the following 
boundary condition at the left boundary $y=-L/2$: 
\bea
\Sigma\left(-\frac{L}{2}\right) &=& 
\diag(m_{A_1},m_{A_2},\ldots, m_{A_{N_c}}), 
 \\
H_{r=A_r}^{A}&=&0, \quad A < A_r. 
\eea
Similarly at the right boundary $y=L/2$, we demand 
\bea
\Sigma\left(\frac{L}{2}\right) &=& 
\diag(m_{B_1},m_{B_2},\ldots, m_{B_{N_c}}), 
 \\
H_{r=B_r}^{A}&=&0, \quad A > B_r. 
\eea
Since Weyl permutations are a part of gauge invariance, 
we need to combine possible Weyl permutations 
of these boundary conditions. 

The BPS equations (\ref{BPS1}), (\ref{BPS2}) and (\ref{BPS3}) 
with the above boundary conditions produce soliton-like 
objects which are localized on the one-dimensional interval 
and connect field configurations specified by the label 
of indices $\vec{A}=(A_1, \cdots, A_{N_c})$ and 
$\vec{B}=(B_1, \cdots, B_{N_c})$.
Since these BPS solitons have unit co-dimension and constructed 
using a non-Abelian gauge theory, 
these BPS solitons are called {\it non-Abelian domain-walls}.

The moduli space of domain-walls is defined by a 
space of parameters of solutions 
of the BPS equations with identification up to 
gauge transformations.
Hence the moduli space is represented by a quotient 
space by the $U(N_c)$ gauge identification 
\be
{\mathcal M}^{N_c,N_f}
_{\vec{A}\to\vec{B}}
 = \frac{\mu_r^{-1}(0)\cap \mu_c^{-1}(0) 
\cap {\mu_c^\dag}^{-1}(0) }{U(N_c)},
\ee
where $\mu_r^{-1}(0)$, $\mu_c^{-1}(0)$ and 
${\mu_c^\dag}^{-1}(0)$ stand for the space of solutions 
of the BPS equations $\mu_r=\mu_c=\mu_c^\dag=0$ with 
the boundary conditions labeled by $\vec{A}$ and $\vec{B}$ 
at $y=-L/2$ and $y=L/2$, respectively. 
This quotient space is known to be a \Kahler quotient space, 
and $\mu_r$, $\mu_c$ and $\mu_c^\dag$ are called moment maps 
in this sense.

The volume of the moduli space is usually defined by an 
integral of the volume form over the whole moduli space 
with $2n$-dimensional coordinates $x$ 
\be
\Volume\left({\mathcal M}^{N_c,N_f}_{\vec{A}\to\vec{B}}\right)
=\int_{{\mathcal M}^{N_c,N_f}_{\vec{A}\to\vec{B}}} 
d^{2n} x \sqrt{\det g_{ij}},
\ee
if we know a metric of the moduli space $g_{ij}$. 
However it is difficult to find the metric of the 
moduli space explicitly in general.

To avoid a direct integration of the volume form on 
the moduli space, we note that the \Kahler manifold admits 
the \Kahler form $\Omega$ and 
the volume form on the \Kahler quotient space can be written 
in terms of $\Omega$ as 
 $d^{2n} x \sqrt{\det g_{ij}} =\frac{1}{n!}\Omega^n$.
On the moduli space, the volume is expressed by 
\be
\Volume\left({\mathcal M}^{N_c,N_f}_{\vec{A}\to\vec{B}}\right)
=\int_{{\mathcal M}^{N_c,N_f}_{\vec{A}\to\vec{B}}} e^\Omega,
\label{volume integral}
\ee
under the consent that the integral exists only on 
the $2n$-form. 

We can also express the volume integral 
(\ref{volume integral}) 
by a path integral over all field configurations 
with suitable constraints onto the moduli space 
${\mathcal M}^{N_c,N_f}_{\vec{A}\to\vec{B}}$ 
\be
\Volume\left({\mathcal M}^{N_c,N_f}_{\vec{A}\to\vec{B}}\right)
=\frac{1}{\Volume(\mathcal{G})}\int
\D \Phi \,
\D \vec{\B}_v \,
\D \vec{\F}_v \,
\D^2 \vec{\B}_m \,
\D^2 \vec{\F}_m  \,
e^{-S_0},
\label{volume of moduli}
\ee
where $\vec{\B}_v=(A_y,\Sigma)$ and 
$\vec{\F}_v=(\lambda_y,\xi)$ are 
vectors of bosonic and fermonic fields in 
the adjoint representation,
$\vec{\B}_m=(H,Y_c)$ and $\vec{\F}_m=(\psi,\chi_c)$ 
are vectors of bosonic and fermonic fields in the 
fundamental representation, and $\Volume(\mathcal{G})$ is the 
volume of $U(N_c)$ gauge transformation group $\mathcal{G}$. 
Precisely speaking, the definition of the volume 
of the moduli space via the path integral 
has an ambiguity corresponding to an ambiguity 
in the definition of the normalization of the metric 
(\Kahler form) of the moduli space. 
We will discuss this point later.

We choose an ``action'' $S_0$ to give constraints on 
the moduli space, which are achieved by integrating 
over Lagrange multiplier fields $\Phi$, $Y_c$ and 
$Y_c^\dag$, and introduce fermions $\lambda_y$, 
$\xi$, $\psi$, $\psi^\dag$, $\chi_c$ and $\chi_c^\dag$ 
to give a suitable \Kahler form on the moduli space 
and Jacobians for the constraints. 
Inspired by the general discussion in \cite{Miyake:2011yr},  
we take the following action $S_0$ 
\begin{multline}
S_0 = i \beta \int_{-\frac{L}{2}}^{\frac{L}{2}} 
dy \, \Tr\Big[
\Phi \mu_r - \lambda_y \xi +i\frac{g^2}{2}\psi\psi^\dag
+ Y_c^\dag \mu_c\\
 -\chi_c^\dag\left(
\frac{\delta \mu_c}{\delta A_y}\lambda_y
+\frac{\delta \mu_c}{\delta \Sigma}\xi
+\frac{\delta \mu_c}{\delta H}\psi
\right)
 + (\text{h.c.})
\Big],
\end{multline}
in order to impose the constraints, and to give the \Kahler 
form and Jacobians in the path integral over the field configurations. 
We also introduced a parameter $\beta$ with a dimension of length. 
Thus the volume of the domain-wall moduli space is evaluated 
by the path integral over fields like a partition 
function of a gauge field theory. 
The role of the Lagrange multiplier field $\Phi$ is 
rather special compared to other fields.
We treat the path integral over 
$\Phi$ separately from other fields.

We can evaluate the integral (\ref{volume of moduli}) 
directly by using a usual field theoretical procedure 
as performed in \cite{Miyake:2011yr}.
However, once we noticed that the action $S_0$ possesses 
an extra symmetry (BRST symmetry), we can evaluate the 
path integral (\ref{volume of moduli}) via the so-called 
localization method (cohomological field theory) 
much more easily than the direct evaluation of the path 
integral. 
We will see that the path integral (\ref{volume of moduli}) 
is localized at the fixed point sets of the BRST symmetry 
and is reduced to a finite dimensional integral.

\section{Localization in Field Theory}

To proceed the evaluation of the path integral 
(\ref{volume of moduli}), we introduce the following 
fermionic transformations (BRST transformations) for the 
vector fields (fields in the adjoint representation)
\be
\begin{array}{ll}
Q A_y = \lambda_y, & Q \lambda_y = -\D_{y}\Phi,\\
Q \Sigma = \xi,& Q \xi = i[\Phi,\Sigma],\\ 
Q \Phi =0,
\end{array}
\label{BRSTvec}
\ee
and for the matter fields (fields in the fundamental representation)
 \be
\begin{array}{ll}
Q H = \psi, &
Q \psi =  i\Phi H,\\
Q Y_c = i  \Phi\chi_c,&
Q \chi_c = Y_c,
\end{array}
\label{BRSTchi}
\ee
and for their hermitian conjugates. 
We see a square of this transformation generates a gauge 
transformation $\delta_G(\Phi)$ with $\Phi$ as the 
transformation parameter: 
 $Q^2 = \delta_G(\Phi)$.
This means that $Q^2$ is nilpotent on gauge invariant 
operators $\Op$. 
If we restrict a space of operator fields to the gauge 
invariant ones, $Q$ gives a cohomology by an identification 
\be
 \Op \sim \Op + Q(\text{gauge inv.~op.}),
\ee
which is called the {\it equivariant cohomology}.

Under this transformation, we find that the action $S_0$ 
is invariant ($Q$-closed) 
\be
Q S_0 = 0.
\ee
We also find that the action $S_0$ can be written by
\begin{multline}
S_0 = i \beta \int_{-\frac{L}{2}}^{\frac{L}{2}} dy \, \Tr\left[
\Phi \left(\D_y\Sigma -\frac{g^2c}{2}{\bf 1}_{N_c}\right)
-\lambda_y \xi\right]\\
+ i \beta Q 
 \int_{-\frac{L}{2}}^{\frac{L}{2}} dy \, \Tr\left[-\frac{g^2}{2}\psi H^\dag
+ \mu_c \chi_c^\dag + \chi_c \mu_c^\dag
\right].
\end{multline}
Here we imposed the periodic boundary condition for the product $\Phi \xi$ in order to preserve the BRST invariance for the action. 
So an essential cohomological part ($Q$-closed but not $Q$-exact) of the action $S_0$ is
\be
S_\text{coh} = i \beta \int_{-\frac{L}{2}}^{\frac{L}{2}} dy \, \Tr\left[
\Phi \left(\D_y\Sigma -\frac{g^2c}{2}{\bf 1}_{N_c}\right)
-\lambda_y \xi\right],
\label{Qclosedaction}
\ee
in terms of the equivariant cohomology.

Using the nature of the BRST symmetry,
we can add an extra $Q$-exact action $Q\Xi$ to $S_0$ without changing 
the path integral, that is, the deformed path integral 
\be
\Volume\left({\mathcal M}^{N_c,N_f}_{\vec{A}\to\vec{B}}\right)
=
\frac{1}{\mathrm{Vol} (\mathcal{G})}
\int 
\D \Phi \,
\D \vec{\B}_v \,
\D \vec{\F}_v \,
\D^2 \vec{\B}_m \,
\D^2 \vec{\F}_m  \,
e^{-S_0-t Q\Xi},
\label{deformed partition function}
\ee
is independent of a deformation parameter (coupling) $t$
since the path integral measure is constructed to be 
$Q$-invariant.
In the $t\to 0$ limit, the path integral 
(\ref{deformed partition function}) reduces to the original 
one which gives the volume of the moduli space. 
When we choose the deformation parameter $t$ appropriately, 
we can evaluate the path integral exactly. 

To evaluate the path integral, we choose $\Xi$ to be 
the following form
\be
tQ\Xi = t_1 Q\Xi_1 + t_2 Q\Xi_2,
\ee
where
\bea
\Xi_1 &=& \int_{-\frac{L}{2}}^{\frac{L}{2}} dy \, \Tr\left[ \mu_c \chi_c^\dag + \chi_c \mu_c^\dag
\right],\\
\Xi_2 &=& \frac{1}{2}\int_{-\frac{L}{2}}^{\frac{L}{2}} dy \Tr
\left[
\vec{\F}_m \cdot (Q \vec{\F}^\dag_m)
 + 
 \vec{\F}_m^\dag  \cdot (Q \vec{\F}_m)
\right].
\label{bosonic part}
\eea
The former $Q\Xi_1$ is already included in the original action $S_0$ and
gives a $\delta$-functional constraint on $\mu_c=\mu_c^\dag=0$ by integrating out the $Y_c$ and $Y^{\dagger}_c$.
This constraint means that the field configuration must satisfy
\be
\D_y H + \Sigma H - H M=0,
\label{constraint for H}
\ee
for the bosonic field $H$.
The fermionic fields in the fundamental representation
must strictly obey the equation of motion
\bea
&&\D_y \psi + \Sigma \psi - \psi M=0,
\label{constraint for psi}
\\
&&\D_y \chi_c - \chi_c \Sigma + M \chi_c=0.
\label{constraint for chi}
\eea
As we will see later,
the above constraints for the fields in the fundamental representation 
are important to count the number of zero modes of the fields at the localization point.

First of all, we introduce Cartan-Weyl basis $(\mathtt{H}_a, \mathtt{E}_\alpha, \mathtt{E}_{-\alpha})$ of the Lie algebra $\mathfrak{u}(N_c)$
and decompose the fields in the adjoint representation as 
follows:
\bea
\Phi & = & \sum_{a=1}^{N_c} \Phi^a \mathtt{H}_a
+\sum_{\alpha>0} \Phi^\alpha \mathtt{E}_\alpha
+\sum_{\alpha>0} \Phi^{-\alpha} \mathtt{E}_{-\alpha},\\
A_y & = & \sum_{a=1}^{N_c} A_y^a \mathtt{H}_a 
+\sum_{\alpha>0} A_y^\alpha \mathtt{E}_\alpha
+\sum_{\alpha>0} A_y^{-\alpha} \mathtt{E}_{-\alpha},\\
\Sigma & = & \sum_{a=1}^{N_c} \Sigma^a \mathtt{H}_a 
+\sum_{\alpha>0} \Sigma^\alpha \mathtt{E}_\alpha
+\sum_{\alpha>0} \Sigma^{-\alpha} \mathtt{E}_{-\alpha},
\eea
where $\mathtt{H}_a$, $\mathtt{E}_\alpha$ and 
$\mathtt{E}_{-\alpha}$ 
satisfy the following commutation relations 
\bea
&&[\mathtt{H}_a,\mathtt{H}_b]=0,\\
&&[\mathtt{H}_a,\mathtt{E}_\alpha]=\alpha_a \mathtt{E}_\alpha,\\
&&\Tr \mathtt{E}_\alpha \mathtt{E}_\beta = \delta_{\alpha+\beta,0},
\eea
and $\mathtt{E}_{-\alpha} = \mathtt{E}_\alpha^\dag$.

To perform the path integral, we introduce the ghosts $c$ 
and $\bar{c}$ for the diagonal gauge fixing condition 
($\Phi^\alpha=0$). 
The ghosts induce the action
\be
S_\text{ghost} = i \beta \int_{-\frac{L}{2}}^{\frac{L}{2}} dy \, \Tr
c[\Phi,\bar{c}],
\ee
which gives a one-loop determinant
\be
\prod_{\alpha>0} \left(\beta \alpha(\Phi)\right)^{2f^\text{adj}_\alpha}.
\label{ghostloneloop}
\ee
where 
$f^{\text{adj}}_{\alpha}$ represents the degree of freedom for each off-diagonal component of real fermion in one-dimension. 
In this gauge choice, the bosonic term for the $Q$-closed action (\ref{Qclosedaction}) can be written as
\be
\left. S_\text{coh} \right|_{\text{bosonic}} = i \beta \int_{-\frac{L}{2}}^{\frac{L}{2}} dy \, \left[
\sum_{a=1}^{N_c} \Phi^a \left(\partial_y\Sigma^a  -\frac{g^2c}{2}\right)  +i\sum_{\alpha} \alpha (\Phi) A_{y}^{-\alpha} \Sigma^{\alpha}  
\right].
\ee
The path integral over off-diagonal elements $(A^{\alpha}_y, \Sigma^{\alpha})$ leads to the one-loop determinant for bosonic fields in the vector multiplet
\be
\prod_{\alpha>0}|\beta \alpha(\Phi)|^{-2b^\text{adj}_\alpha}.
\label{vecbosoniconeloop}
\ee 
From (\ref{ghostloneloop}) and (\ref{vecbosoniconeloop}), we obtain the one-loop determinant for off-diagonal elements in the adjoint 
representation
\be
\prod_{\alpha>0}
\frac{\left(\beta \alpha(\Phi)\right)^{2f^\text{adj}_\alpha}}
{ |\beta \alpha(\Phi) |^{2b^\text{adj}_\alpha}}.
\ee

Naively scalar fields and vector fields carry the same 
degrees of freedom in one-dimension, so we can conclude 
that $b^\text{adj}_\alpha = f^\text{adj}_\alpha$, 
that is, the one-loop determinants for the adjoint fields 
are canceled out up to a signature $\pm 1$.
It is difficult to determine the signature of the one-loop 
determinant at this stage, but we will assume later that this 
signature depends on permutations of the boundary conditions.
We can non-trivially check that this assumption is consistent and leads correct answers
to the volume and dualities of the domain-walls.

Next we evaluate the one-loop determinant of the fields in 
the fundamental representation. The matter fields 
enter in the action through the $Q$-exact term;
\bea
Q \Xi_2 &=& \int_{-\frac{L}{2}}^{\frac{L}{2}} dy 
\left[ i \sum_{a=1}^{N_c} (H^{\dagger}_a \Phi^a H_a 
+ \chi^{\dagger}_{c, a} \Phi^a \chi_{c, a} ) 
+ i \psi^{\dagger} \psi + i Y^{\dagger}_c Y_c  \right].
\eea
The matter action is quadratic with respect to the field 
in the fundamental representation, so we can perform the 
path integral and obtain the one-loop determinant;
\be
\prod_{a=1}^{N_c} (i \Phi^a)^{f^\text{fund}_a-b^\text{fund}_a }.
\label{eq:1loop-fundmental}
\ee
Here $f^\text{fund}_a$ and $b^\text{fund}_a$ are the degrees 
of freedom for the fundamental fields $\chi_{c, a}$ and 
$H_a$, respectively.

On the other hand, since fields in the fundamental 
representation originally obey the constraints 
 (\ref{constraint for H}) -- (\ref{constraint for chi}), 
when we define the differential operator for the general fields $\Psi_a$ and $\tilde{\Psi}_a$
in the fundamental representation 
by
\be
P_a \Psi_a \equiv \D_y \Psi_a+\Sigma \Psi_a-\Psi_a M, 
\ee
and
\be
\tilde{P}_a \tilde{\Psi}_a \equiv \D_y \tilde{\Psi}_a
-\Sigma \tilde{\Psi}_a+\tilde{\Psi}_a M, 
\ee 
the fields $(H_a, H^{\dagger}_{a})$ and 
$(\chi_{c,a}, \chi^{\dagger}_{c,a})$ should be expanded by 
the eigenmodes for the operators $P_a$ and $\tilde{P}_a$, 
respectively. 
Since $P_a$ and $\tilde{P}_a$ are adjoint for each 
other, their eigenmodes coincide including the degeneracy 
and their difference in Eq.(\ref{eq:1loop-fundmental}) are 
canceled out except for the zero modes. 
Thus we find the difference of the number of the modes for 
the fields in the fundamental representation is characterized 
by the dimensions of zero modes, i.e. the index 
\begin{equation}
\ind P_a \equiv  \dim \ker P_a - \dim \ker \tilde{P}_a,
\label{eq:index}
\end{equation}
The one-loop determinant of the matter fields becomes
\be
\prod_{a=1}^{N_c} \frac{1}{(i\Phi^a)^{\ind P_a}}.
\ee
Note that the index of $P_a$     
will depends only on the boundary condition of $\Sigma^a$ 
similarly to the Callias index theorem \cite{Callias:1977kg}.
We will show how to compute this index for various examples 
in the next section.

Thus the path integral (\ref{deformed partition function}) 
reduces to that of a direct product $U(1)^{N_c}$ of Abelian 
gauge theories after the off-diagonal components of the 
fields are integrating out 
\be
\Volume\left({\mathcal M}^{N_c,N_f}_{\vec{A}\to\vec{B}}\right)
=\frac{1}{\mathrm{Vol}(\mathcal{H})N_c! }
\prod_{a=1}^{N_c} 
\int 
\D \Phi^a
\D A_y^a
\D \Sigma^a
\D \lambda^a_y
\D \xi^a
\frac{1}{(i\Phi^a)^{\ind P_a}}
e^{- S^a_\text{coh}[\Phi^a,\Sigma^a]},
\label{abelian theory}
\ee
where
\be
S^a_\text{coh}[\Phi^a,\Sigma^a]
=i \beta  \int_{-\frac{L}{2}}^{\frac{L}{2}} dy \, 
\left[ \Phi^a\left(
\del_y \Sigma^a-\frac{g^2c}{2}
\right)-\lambda_y^a \xi^a
\right],
\label{abelian action}
\ee
and $\mathrm{Vol}(\mathcal{H})$ is the volume of 
the gauge transformation group of  $\mathcal{H}=U(1)^{N_c}$. 
The pre-factor $1/N_c!$ comes from the order of the Weyl 
permutation group in the original $U(N_c)$ gauge group. 

To perform the path integral (\ref{abelian theory}) of the 
$U(1)^{N_c}$ gauge theory, we choose a gauge $A_y^a=0$ and 
expand $\Sigma^a$ around a specific profile function 
$\Sigma_0^a$ by 
\be
\Sigma^a(y) = \Sigma^a_0(y) + \tilde{\Sigma}^a(y),
\ee
where $\Sigma^a_0$ satisfies the given boundary condition 
 at $y=-\frac{L}{2}$ and $y=\frac{L}{2}$.
We note that there still exists a degree of freedom of
the Weyl permutation group after fixing the gauge and 
the ``classical'' background profile $\Sigma^a_0$ 
satisfying the boundary condition.

A partial integration over the fluctuations 
$\tilde{\Sigma}^a$ 
of the action (\ref{abelian action}) gives the constraint 
$\del_y \Phi^a=0$ 
as expected from the localization.
So the path integral over $\Phi^a(y)$ reduces to an 
integration over constant modes\footnote{
We use a subscript of the Cartan indices $a$ for the 
constant modes 
for the later appearance of equations.
} $\phi_a$.

In the original non-Abelian gauge theory, the boundary 
condition is chosen to be 
$\Sigma(-\frac{L}{2})=\diag(m_{A_1},m_{A_2},\ldots,m_{A_{N_c}})$ and
$\Sigma(\frac{L}{2})=\diag(m_{B_1},m_{B_2},\ldots,m_{B_{N_c}})$
up to the Weyl permutations. 
For a given Weyl permutation $\sigma$, the boundary condition 
for the background profile $\Sigma_0^a$ becomes 
$\Sigma_0^a(-\frac{L}{2})=m_{A_{\sigma(a)}}$ and 
$\Sigma_0^a(\frac{L}{2})=m_{B_{\tilde\sigma(a)}}$, 
where $\sigma(a)$ and $\tilde\sigma(a)$ are elements of 
the permutation group $\mathfrak{S}_{N_c}$.
The above choice of the boundary condition gives 
the classical value of the action at the fixed points as 
\bea
S^a_\text{coh}[\phi_a,\Sigma_0^a]
&=&i \beta \phi_a \int_{-\frac{L}{2}}^{\frac{L}{2}} dy
\left\{
\del_y \Sigma_0^a-\frac{g^2c}{2}
\right\}\nn\\
&=&i \beta \phi_a \left\{
(m_{B_{\tilde\sigma(a)}}-m_{A_{\sigma(a)}})-\frac{g^2c}{2}L
\right\},
\eea
for the permutation $\sigma$ and $\tilde\sigma$.

Using this evaluation of the cohomological action at the 
fixed points, 
 we  obtain the integral formula for the volume of moduli 
space of the domain-walls after integrating out all of 
fluctuations of the fields
\be
\Volume\left({\mathcal M}^{N_c,N_f}_{\vec{A}\to\vec{B}}\right)
=\frac{1}{N_c!}
\sum_{(\sigma,\tilde\sigma)\in(\mathfrak{S}_{N_c})^2} 
\prod_{a=1}^{N_c} 
\int_{-\infty}^\infty  
\frac{d \phi_a}{2\pi}
\frac{(-1)^{|\sigma||\tilde\sigma|}}{(i\phi_a)^{\ind P_a}}
e^{i\beta\phi_a \left\{
\hat{L}-(m_{B_{\tilde\sigma(a)}}-m_{A_{\sigma(a)}})
\right\}},
\label{eq:double-sum}
\ee
where we define 
\be
\hat{L}\equiv \frac{g^2c}{2}L, 
\label{eq:def-hat-L}
\ee
and introduce the signature dependence which is determined 
by the order of the permutations $|\sigma|$ 
and $|\tilde\sigma|$.
As explained before, the signature dependence coming from 
the one-loop determinant is not obvious, but we will see 
that this assumption works well and pass non-trivial checks 
in the later discussions. 

Since (\ref{eq:double-sum}) depends only on the relative 
permutation between $\sigma$ and $\tilde\sigma$, 
a sum over one permutation simply cancels $1/{N_c}!$ and 
only a sum over the relative permutation remains 
\be
\Volume\left({\mathcal M}^{N_c,N_f}_{\vec{A}\to\vec{B}}\right)
=\sum_{\sigma\in \mathfrak{S}_{N_c}} \prod_{a=1}^{N_c}
\int_{-\infty}^\infty  
\frac{d \phi_a}{2\pi}
\frac{(-1)^{|\sigma|}}{(i\phi_a)^{\ind P_a}}
e^{i\beta\phi_a \left\{
\hat{L}-(m_{B_{\sigma(a)}}-m_{A_a})
\right\}}.
\label{integral formula}
\ee
We will apply this formula, 
which is written by an integral over the constant modes
of $\Phi^a$ and a summation over the Weyl permutation group 
of the boundary conditions,
 to evaluate explicitly various examples of domain-walls 
in the next section.

\section{Various Examples}

\subsection{Abelian domain-walls}

In this section, we give some examples of the volume of 
the moduli space of domain-walls following the general 
formula (\ref{integral formula}). 
A key to evaluate the volume concretely is a computation 
of the index of the operator $P_a$.
We will see the index is obtained from (topological) 
profile of the function $\Sigma(y)$.

We first show how to evaluate the volume of the domain-wall 
moduli space for Abelian gauge theories. 
The integral formula (\ref{integral formula}) for non-Abelian 
gauge theories is essentially a direct product of Abelian 
gauge theories, except for the existence of the permutations, 
thanks to the localization. 
Then if we obtain the volume of moduli space of the Abelian 
domain-walls, we can easily extend it to the non-Abelian case.
So we here would like to explain carefully a detail of 
the Abelian case.

To make an example more explicit, we consider the case 
$N_c=1$ (Abelian) and 4 flavors $N_f=4$. 
The mass for $H$ and $H^\dag$ can be set 
$M=\diag(m_1,m_2,m_3,m_4)$ with $m_1<m_2<m_3<m_4$ 
without loss of generality. 
We also impose the boundary condition 
$\Sigma(-\frac{L}{2}) =m_1$ and $\Sigma(\frac{L}{2})=m_4$ 
as the first example.

Applying the integral formula in Eq.(\ref{integral formula}) 
to the case of $N_c=1$ and $N_f=4$, we obtain for 
this example, 
\be
\Volume\left({\mathcal M}^{1,4}_{1 \to 4}\right)
=
\int_{-\infty}^\infty  
\frac{d \phi}{2\pi}
\frac{1}{(i\phi)^{\ind P}}
e^{i\phi \beta \left(
\hat{L}-(m_4-m_1)
\right)}, 
\ee
where we suppressed the suffix $a$ in $\phi_a$, $P_a$ 
and so on, since $a=1$ for the $N_c=1$ case. 
To perform this integral, we have to determine the 
index of $P$ defined in Eq.(\ref{eq:index}).

\subsubsection{Counting of zero modes}

Let us first consider a differential equation
\be
P \Psi_i = \del_y \Psi_i +\sum_{j=1}^4 A_{ij}(y)\Psi_j = 0,
\label{differential equation 1}
\ee
where $A_{ij}(y)\equiv (\Sigma(y)-m_i)\delta_{ij}$.
We define the kernel of $P$ as ``normalizable'' modes of 
the solution of the above differential equation $\Psi_i(y)$.
Although a term ``normalizable'' is used here, it is 
actually not determined by the convergent normalization 
of the mode function, but is determined by physical 
considerations as described below. 
We will also give a mathematically more precise definition 
later.

In order to find these normalizable or non-normalizable modes 
concretely,
let us assume simply that a profile of $\Sigma(y)$ is a 
straight line 
\be
\Sigma(y) = \frac{d}{L}y+\overline{m},
\ee
where $d\equiv m_4-m_1$ and $\overline{m}\equiv \frac{m_1+m_4}{2}$.
Using this profile, we can solve the differential equation
(\ref{differential equation 1}) and (\ref{differential equation 2}).
The result is 
\bea
\Psi(y)&=&\left(
C_1e^{-\frac{d}{2L}(y+\frac{L}{2})^2},
C_2e^{-\frac{d}{2L}\left(y-\frac{L(m_2-\overline{m})}{d}\right)^2},
C_3e^{-\frac{d}{2L}\left(y-\frac{L(m_3-\overline{m})}{d}\right)^2},
C_4e^{-\frac{d}{2L}(y-\frac{L}{2})^2}
\right), \nn\\
\tilde{\Psi}(y)&=&\left(
\tilde C_1e^{\frac{d}{2L}(y+\frac{L}{2})^2},
\tilde C_2e^{\frac{d}{2L}\left(y-\frac{L(m_2-\overline{m})}{d}\right)^2},
\tilde C_3e^{\frac{d}{2L}\left(y-\frac{L(m_3-\overline{m})}{d}\right)^2},
\tilde C_4e^{\frac{d}{2L}(y-\frac{L}{2})^2}
\right),\nn\\
\eea
with integration constants $C_i, \tilde C_i, i=1,\cdots,4$. 
Since $d>0$, all the solutions of $\tilde{\Psi}$ rapidly
diverge at the boundary of the interval when  $L$ is sufficiently large.
We call these divergent modes for the large $L$ as ``non-normalizable''.
On the other hand, the functions in $\Psi(y)$ are Gaussian and damp well
at the boundary.
We classify 
these modes are ``normalizable''.
The number of normalizable modes is four in $\Psi$ 
for any size $L$ of the interval.
These observations imply that $\dim \ker P = 4 $ and 
$\dim \ker \tilde{P}=0$.
So we find $\ind P = \dim \ker P - \dim \ker \tilde{P} = +4$.

We need to be careful when we consider other boundary 
conditions where the profile of $\Sigma(y)$ does not 
reach some of the values of masses. 
For instance, 
if we consider a different boundary condition 
$\Sigma(-\frac{L}{2})=m_2$ 
and $\Sigma(\frac{L}{2})=m_4$, namely the profile of $\Sigma(y)$
does not reach at $\Sigma=m_1$ and $A_{11}(y)=\Sigma(y)-m_1$ is always positive.
In this case, the function $\Psi_1(y)$ behaves as
\be
\Psi_1(y) = C  \, e^{-\frac{d'}{2L}\left(y
+\frac{L(\overline{m}'-m_1)}{d'}\right)^2},
\label{outside function}
\ee
where $d'=m_4-m_2$, $\overline{m}'=(m_2+m_4)/2$, 
and $C$ is an integration constant. 
This mode should be normalizable in the previous sense 
that the function damps 
at the boundaries for large $L$. 
However this kind of functions, which is 
monotonously decreasing or increasing in the interval, 
is localized outside of the interval 
since there is no zeros of $A_{11}(y)$ within the interval.
The localized position of $\Psi_i(y)$ corresponds to 
the position moduli of walls.
We should not include the position moduli made outside of the 
interval.
So we exclude the localized modes expressed like 
(\ref{outside function})
by setting $C=0$ as the boundary condition.

More generally, the signature of the function 
$\Sigma(y)-m_i$ can change between 
$y=-\frac{L}{2}$ and $y=\frac{L}{2}$. 
When the signature of $\Sigma(y)-m_i$ changes from negative 
to positive, a new normalizable mode 
appears for $P$. 
Since we have chosen the boundary condition as 
$\Sigma(y)-m_i=0$ ($i=1,4$), the signature change at the 
boundary is a little ambiguous. 
We regard the contribution of the signature change from $0$ 
to positive as the same as the change from negative to 
positive. Namely we assume the existence of the 
function $\Sigma(y)$ outside of the interval. 

The kernel of $\tilde{P}$ is also evaluated in a similar 
way as $\ker P$. 
The differential equation for $\tilde{P}$ is now given by 
\be
\tilde{P}\tilde{\Psi}_i = \del_y \tilde{\Psi}_i 
- \sum_{j=1}^4 A_{ij}(y)\tilde{\Psi}_j = 0.
\label{differential equation 2}
\ee
Since the signature in front of the matrix operator 
$A_{ij}= (\Sigma(y)-m_i)\delta_{ij}$ is opposite to the $P$ 
case, the counting of the normalizable modes is completely 
opposite. 
The normalizable modes come from the change of the signature 
of $\Sigma(y)-m_i$ from positive to negative.

\subsubsection{Index theorem}

This counting of the index of $P$, by choosing a specific profile function of $\Sigma$ and
thinking physically whether the mode is normalizable or not,
appears a little bit ambiguous. However we can define 
clearly the index of $P$ in a mathematical way 
which is similar to the 
Atiyah-Patodi-Singer \cite{AtiyahPatodiSinger}
or Callias index theorem \cite{Callias:1977kg}.

The profile of the function $\Sigma(y)$ is completely
determined by the original BPS equations,
especially by solving the equation $\mu_r=0$.
However, in our derivation of the integral formula,
we did not take account of one of the BPS equations 
$\mu_r=0$ before integrating $\phi$.
So while the index is considered for a $P$ with a specific 
$\Sigma(y)$, the index is actually independent of choices 
of the profile of $\Sigma(y)$.

\begin{figure}
\begin{center}
\includegraphics[scale=0.4]{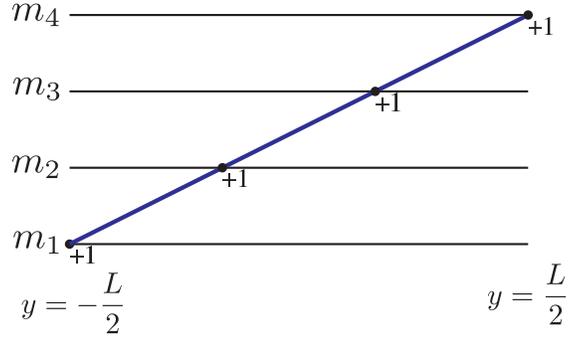}
\end{center}
\caption{The contribution to the index of $P$ when the 
profile of $\Sigma(y)$ is 
a straight line. The straight profile crosses the mass level from negative to positive.
Each crossing contributes to the index by $+1$. So the index is $+4$ in total.}
\label{straight line}
\end{figure}

To see this, let us consider a kink-like profile which may 
be realized by solving the full BPS equations including 
$\mu_r$ to examine the index (\ref{eq:index}) for the $N_f=4$ 
case. 
At the boundary $y=-\frac{L}{2}$,
the eigenvalues of $A_{ij}(y)$ is $(0,+,+,+)$ 
($+$ means a positive eigenvalue).
Since we consider extending the function $\Sigma(y)$ 
infinitesimally outside of the interval $y<-L/2$, 
the eigenvalues at $y=-\epsilon -\frac{L}{2}$ 
is $(-,+,+,+)$. Going through the boundary $y=-\frac{L}{2}$  
we obtain a contribution to the index by $+1$. 
When $\Sigma(y)$ reaches $m_2$ at some $y$, the 
eigenvalues changes 
from $(-,+,+,+)$ to $(-,-,+,+)$, that is, the index 
increases by $+1$.
If we continue to $y=\frac{L}{2}+\epsilon$ in this way, 
we obtain the value of 
the index to be $\ind P=+4$. (See Fig.~\ref{straight line}.)

\begin{figure}
\begin{center}
\begin{tabular}{ccc}
\includegraphics[scale=0.37]{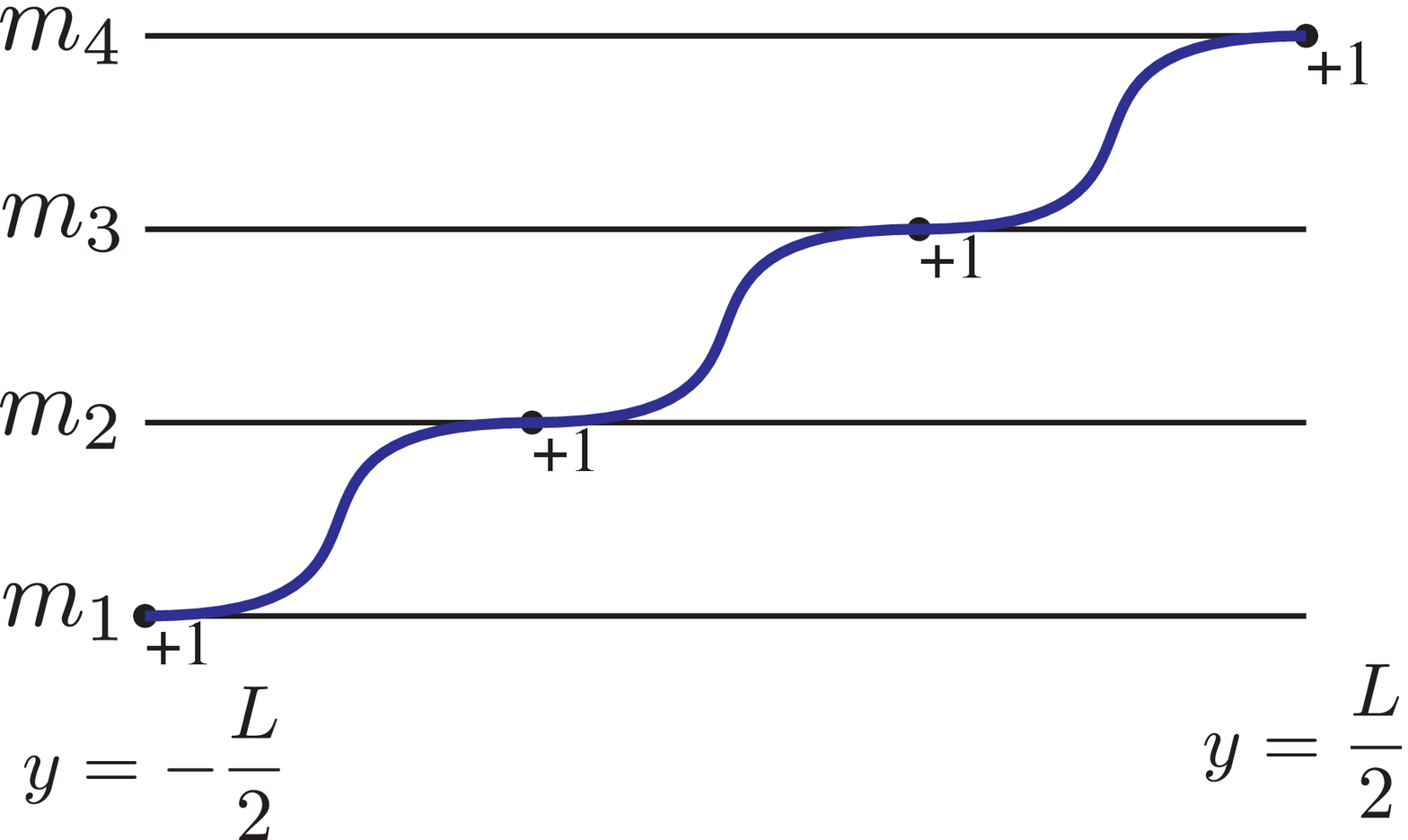} & &
\includegraphics[scale=0.37]{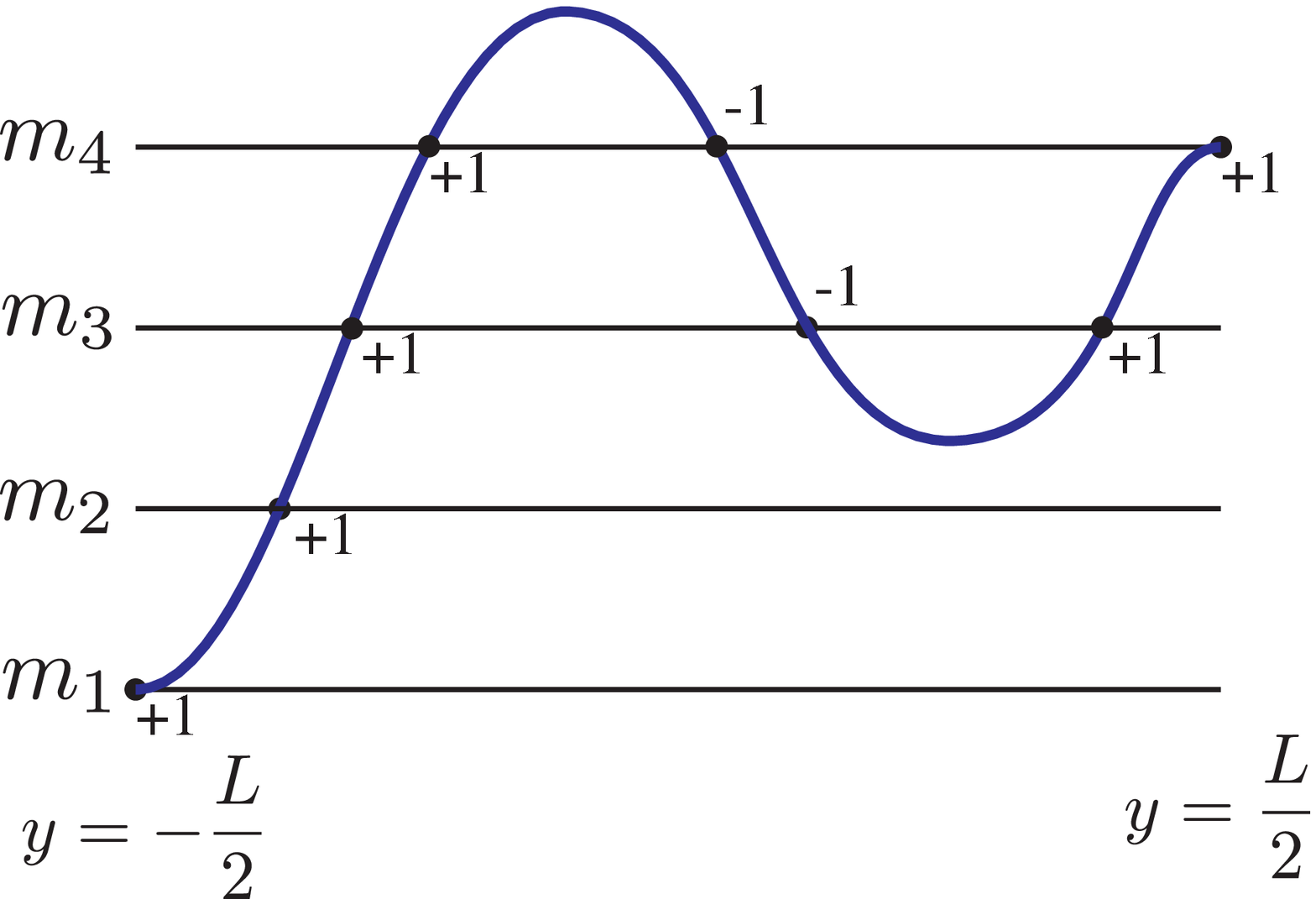} \\
(a) && (b)
\end{tabular}
\end{center}
\caption{The index does not change by continuous 
deformations of the profile of $\Sigma(y)$ with 
the fixed boundary condition. 
(a) The kink profile which may be obtained after solving 
the BPS equations 
 gives the same index as the straight line.
(b) Even if a continuous deformation produces negative 
contributions to the index, additional positive contributions 
are also produced and the total contribution to the index 
remains the same as $+4$.}
\label{generic profile}
\end{figure}

When we choose the profile $\Sigma(y)$ freely, we always 
obtain the same index $\ind P=+4$. 
So the index is invariant under a continuous deformation 
of $\Sigma(y)$. (See Fig.~\ref{generic profile}.)

\subsubsection{Evaluation of integral}

Thus we have the indices for the $N_f=4$ 
case, 
and obtain the integration formula for
the volume of the moduli space as 
\be
\Volume\left({\mathcal M}^{1,4}_{1 \to 4}\right)
=
\int_{-\infty}^\infty  
\frac{d \phi}{2\pi}
\frac{1}{(i\phi)^4}
e^{i\phi \beta \left(
\frac{g^2c}{2}L-d
\right)}.
\label{integral of 1to 4}
\ee
This integral has a fourth order pole at $\phi=0$.
We can perform this integral by using the following 
residue calculus with a suitable contour dictated by the 
convergence of $H, H^\dagger$ path integral 
\be
\int_{-\infty-i\e}^{\infty-i\e}
\frac{d\phi}{2\pi i}
\frac{1}{\phi^{n+1}}
e^{i\phi B}
=
\begin{cases}
\frac{1}{n!}(iB)^n & \text{if $B\geq 0$ and $n\geq0$}\\
0 & \text{otherwise}
\end{cases}.
\label{contour integral}
\ee
Thus we obtain the volume of the moduli space
\be
\Volume\left({\mathcal M}^{1,4}_{1 \to 4}\right)=
\frac{1}{3!}
 \left(
\frac{g^2c}{2}L-d
\right)^3,
\label{volume of 1 to 4}
\ee
when $\frac{g^2c}{2}L-d\geq 0$.

We next discuss implications of the result 
(\ref{volume of 1 to 4}).
If we consider the case $L\gg \frac{2d}{g^2c}$, where 
the size of the interval $L$ is sufficiently large
in comparison with the width 
\cite{Kaplunovsky:1998vt, Eto:2006ng, Shifman:2002jm} 
of the domain-wall $\frac{2d}{g^2c}$,
then the volume is proportional to $\frac{L^3}{3!}$.
This is nothing but
a volume of the moduli space of three undistinguished points on 
the interval $L$.
So we can regard the power of $(\frac{g^2c}{2}L-d)$
as the number of the BPS domain-walls on the interval
(the dimension of the domain-wall moduli space).
This agrees with the number of kinks which is depicted in 
Fig.\ref{generic profile}(a).
Recalling that the order of pole comes from the index of $P$,
so we can conclude that
\be
\ind P = \text{(the number of BPS domain-walls)+1}.
\label{eq:index_wall_number}
\ee

We also understand this fact from another 
point of view.
The index of $P$
is obtained from the equations $\mu_c=\mu_c^\dag=0$
without imposing 
the other BPS equations $\mu_r=0$.
Only after $\phi$ is integrated out, 
the equation $\mu_r=0$ is taken into account.
The number of the domain-walls coincides with the dimension 
of the moduli space. Additional $+1$ of the index of $P$ is 
removed by the contour integral and the number reduces to 
the dimension of the moduli space.
The dimension of the moduli space is also calculated by an 
index theorem where all of the BPS equations are considered.
We have finally obtained the dimension of the moduli space 
after imposing the condition $\mu_r=0$.
Thus we have done a correct evaluation of the moduli space 
volume by the contour integral of $\phi$.

Using similar arguments as above, we can easily extend 
our computation to the case where $N_f$ and the boundary 
conditions are general.
\bea
\Volume\left({\mathcal M}^{1,N_f}_{i \to j}\right)
&=&
\int_{-\infty}^\infty  
\frac{d \phi}{2\pi}
\frac{1}{(i\phi)^{j-i+1}}
e^{i\phi \beta \left(
\frac{g^2c}{2}L-d_{ij}
\right)}\nn\\
&=&
\frac{\beta^{j-i}}{(j-i)!}
\left(
\frac{g^2c}{2}L-d_{ij}
\right)^{j-i},
\eea
where $d_{ij}\equiv m_j-m_i$.

\subsection{Non-Abelian domain-walls}

The localization formula (\ref{abelian theory}) 
says that the non-Abelian gauge group $U(N_c)$ 
reduces to a product of the Abelian groups $U(1)^{N_c}$ 
at the fixed point set.
So the localization formula for the non-Abelian gauge group 
is essentially a direct product of the formula for the 
Abelian group. 
In particular, the indices (number of walls) for each 
Abelian factor is determined by the boundary conditions 
as in Eq.(\ref{eq:index_wall_number}). 
With this result for the indices ind$P$, 
Eq.(\ref{integral formula}) for the volume of the moduli 
space of the non-Abelian walls becomes 
\be
\Volume\left({\mathcal M}^{N_c,N_f}_{\vec{A} 
\to \vec{B}}\right)
=
\sum_{\sigma\in\mathfrak{S}_{N_c}}
\prod_{a=1}^{N_c}
\int_{-\infty}^\infty  
\frac{d \phi_a}{2\pi}
\frac{(-1)^{|\sigma|}}{(i\phi_a)^{B_{\sigma(a)}-A_a+1}}
e^{i\phi_a \beta \left(
\hat{L}-(m_{B_{\sigma(a)}}-m_{A_a})
\right)}.
\label{non-Abelian residue integral}
\ee

If some of the permutations of the boundary conditions satisfy 
$B_{\sigma(a)}-A_a <0$, the corresponding $\phi_a$ integral 
does not have a pole and vanishes. 
These boundary conditions $B_{\sigma(a)}-A_a <0$ correspond 
to non-BPS wall solutions.
Although the non-BPS walls are in general contained 
in the $\phi_a$ integral 
(\ref{non-Abelian residue integral}), they give vanishing 
contributions. 
So the integral is finally restricted to 
a set of permutations $\mathfrak{S}'_{N_c}$,
which satisfy  $\forall (B_{\sigma(a)}-A_a)\geq 0$
(BPS wall conditions). The integral can be evaluated by
\be
\Volume\left({\mathcal M}^{N_c,N_f}_{\vec{A} 
\to \vec{B}}\right)
=
\beta^D
\sum_{\sigma\in\mathfrak{S}'_{N_c}}
\prod_{a=1}^{N_c}
\frac{(-1)^{|\sigma|}}{(B_{\sigma(a)}-A_a)!}
\left(
\hat{L}-(m_{B_{\sigma(a)}}-m_{A_a})
\right)^{B_{\sigma(a)}-A_a},
\label{eq:nonabelian_volume_formula}
\ee
where
$D=\dim {\mathcal M}^{N_c,N_f}_{\vec{A} \to \vec{B}}
=\sum_{a=1}^{N_c}(B_a-A_a)$
is the dimension of the moduli space.

It is interesting to note that the above volume formula 
of the non-Abelian domain-wall 
can be expressed by a determinant of a matrix ${\mathcal T}$ 
\be
\Volume\left({\mathcal M}^{N_c,N_f}_{\vec{A} 
\to \vec{B}}\right)
=\beta^D \det {\mathcal T}^{N_c,N_f}_{\vec{A} \to \vec{B}},
\label{eq:volume_transition_matrix}
\ee
where
\be
\left({\mathcal T}^{N_c,N_f}_{\vec{A} \to \vec{B}}\right)_{ab}
=
\left\{
\begin{array}{ll}
{\displaystyle \frac{1}{(B_b-A_a)!}\left(
\hat{L}-(m_{B_b}-m_{A_a})
\right)^{B_b-A_a}}
 & \text{if } B_b\geq A_a,\\
0 &  \text{if } B_b < A_a.
\end{array}
\right.
\label{eq:def_transition_matrix}
\ee
We will call this matrix ${\mathcal T}$ as a transition 
matrix in the following.

\begin{figure}
\begin{center}
\begin{tabular}{ccc}
\includegraphics[scale=0.37]{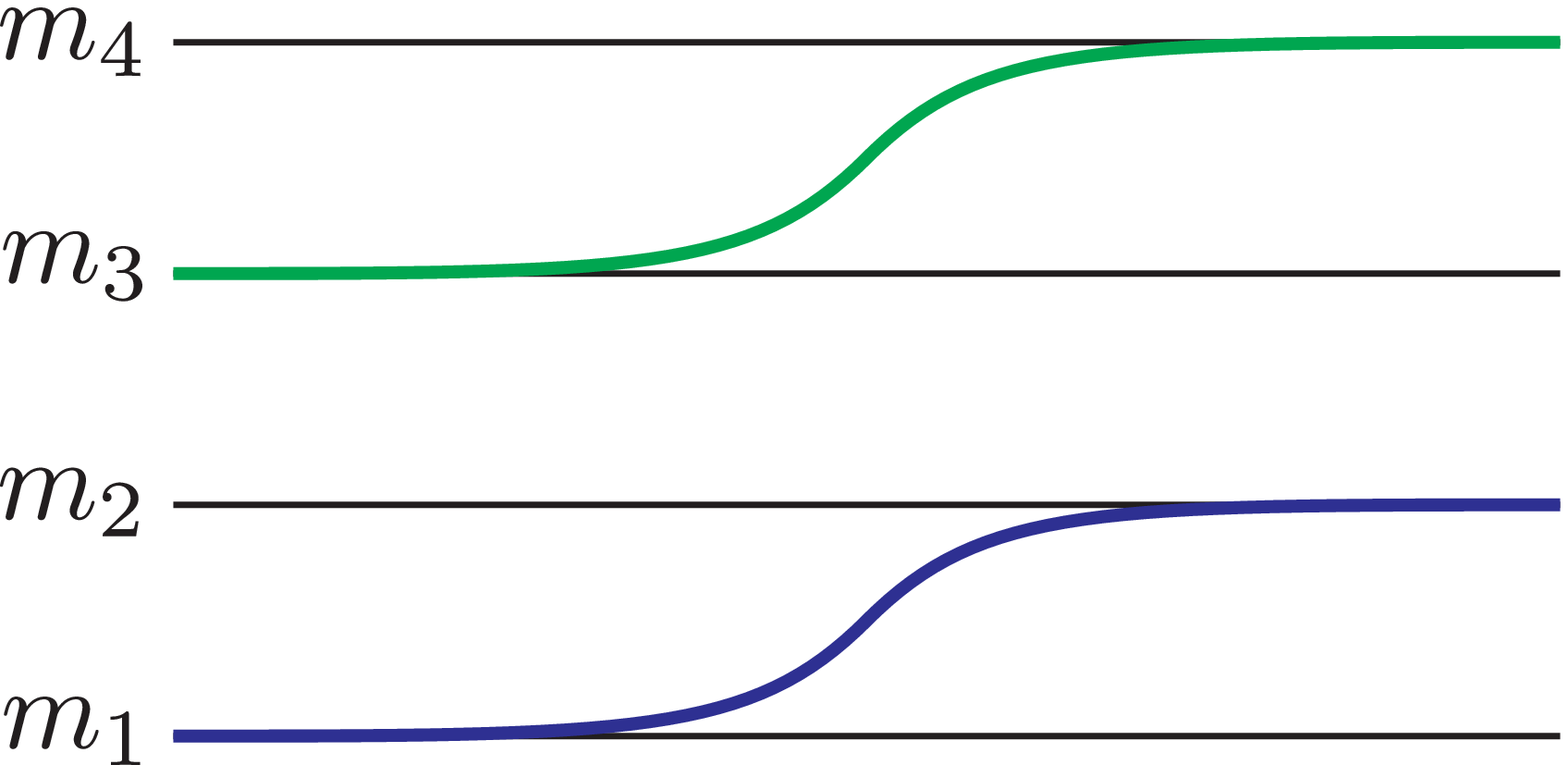} & 
\raisebox{1.4cm}{-} &
\includegraphics[scale=0.37]{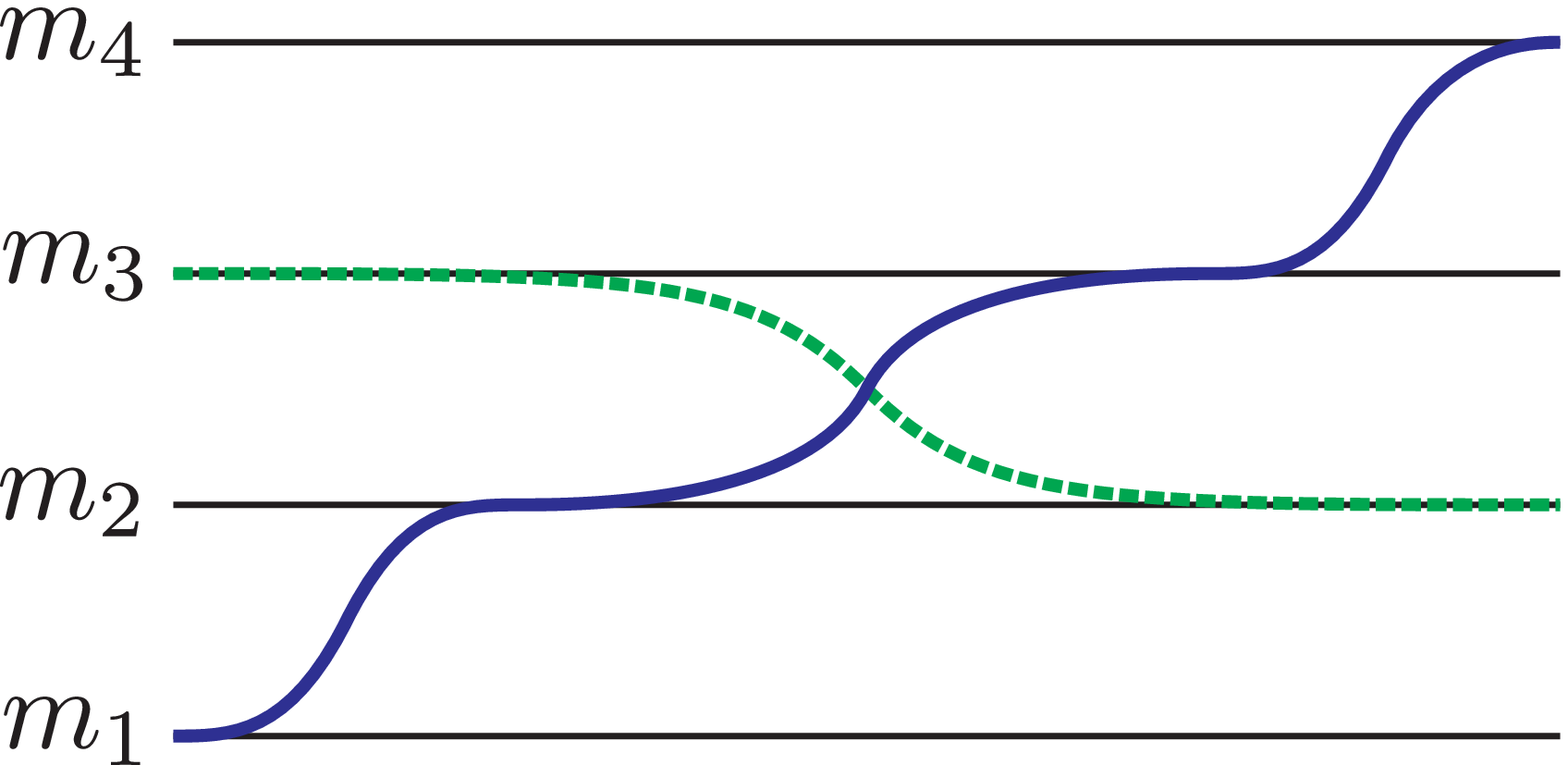}\\
(a) && (b)
\end{tabular}
\end{center}
\caption{The possible domain-wall profiles of the 
boundary condition $\Sigma(-L/2)=\diag(m_1,m_3)$ and 
 $\Sigma(L/2)=\diag(m_2,m_4)$ with the permutations. 
Solid line represents for the BPS profile while dashed 
line represents the anti-BPS one. 
The contribution from the non-BPS domain-wall profile 
which includes the anti-BPS domain-wall disappears from 
the sum of the volume evaluation.}
\label{NAW 13to24}
\end{figure}

Using the above formula, let us consider some concrete 
examples for the non-Abelian gauge group 
in order to understand the meaning of the volume formula 
(\ref{eq:nonabelian_volume_formula}).
We first consider 
the case of $N_c=2$ and $N_f=4$ with
the boundary condition $\Sigma(-L/2)=\diag(m_1,m_3)$ and
 $\Sigma(L/2)=\diag(m_2,m_4)$.
 The $\phi_a$ integral (\ref{eq:nonabelian_volume_formula}) 
for this boundary condition is 
given concretely  by
 \bea
 \Volume\left({\mathcal M}^{2,4}_{(1,3) \to (2,4)}\right)
& =&
 \int \frac{d\phi_1}{2\pi}\frac{d\phi_2}{2\pi}
 \frac{ e^{i\beta \phi_1(\hat{L}-(m_2-m_1))}}{(i\phi_1)^2}
\frac{e^{i\beta \phi_2(\hat{L}-(m_4-m_3))}
}{(i\phi_2)^2}\nn\\
&&
 - \int \frac{d\phi_1}{2\pi}\frac{d\phi_2}{2\pi}
 \frac{ e^{i\beta \phi_1(\hat{L}-(m_4-m_1))}}{(i\phi_1)^4}
\frac{e^{i\beta \phi_2(\hat{L}-(m_2-m_3))}
}{(i\phi_2)^{0}}.
 \eea
The second term contains the anti-BPS wall configuration and 
the $\phi_a$ integral vanishes. 
So only the first term contributes to the volume. 
Thus we obtain
\be
\Volume\left({\mathcal M}^{2,4}_{(1,3) \to (2,4)}\right)
=\beta^2
(\hat{L}-(m_2-m_1))
(\hat{L}-(m_4-m_3)).
\label{Vol of NAW 13to24}
\ee
This result is essentially a direct product of 
independent Abelian walls. (See Fig.~\ref{NAW 13to24}.)
In this case, the $2\times2$ transition matrix 
 becomes 
\be
{\mathcal T}^{2,4}_{(1,3) \to (2,4)}
=
\begin{pmatrix}
\hat{L}-(m_2-m_1) & \frac{1}{3!}(\hat{L}-(m_4-m_1))^3 \\
0 & \hat{L}-(m_2-m_1)
\end{pmatrix}.
\ee
The determinant of this matrix (times $\beta^2$) 
precisely gives (\ref{Vol of NAW 13to24}).

\begin{figure}
\begin{center}
\begin{tabular}{ccc}
\includegraphics[scale=0.37]{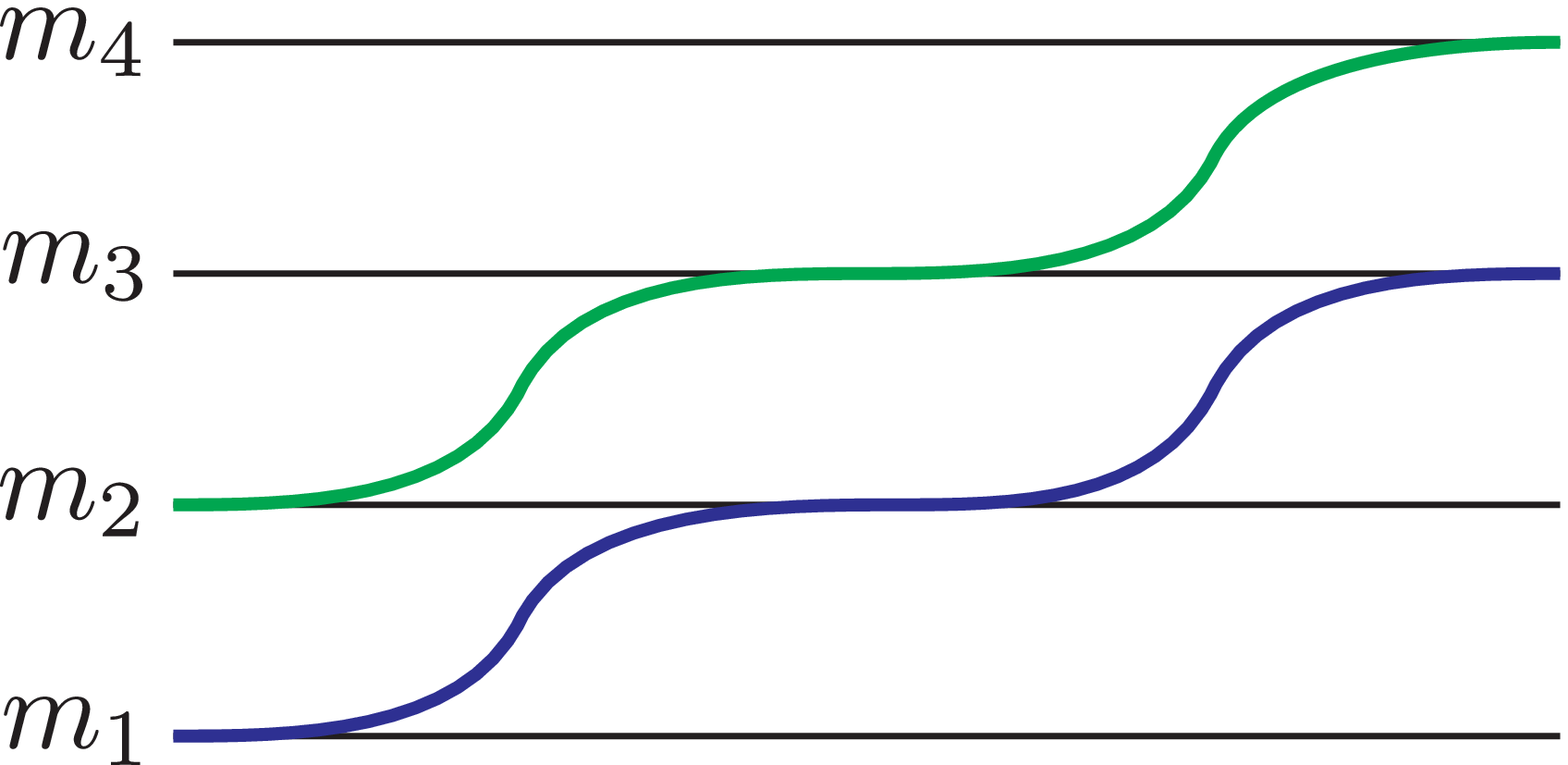} & \raisebox{1.4cm}{-} &
\includegraphics[scale=0.37]{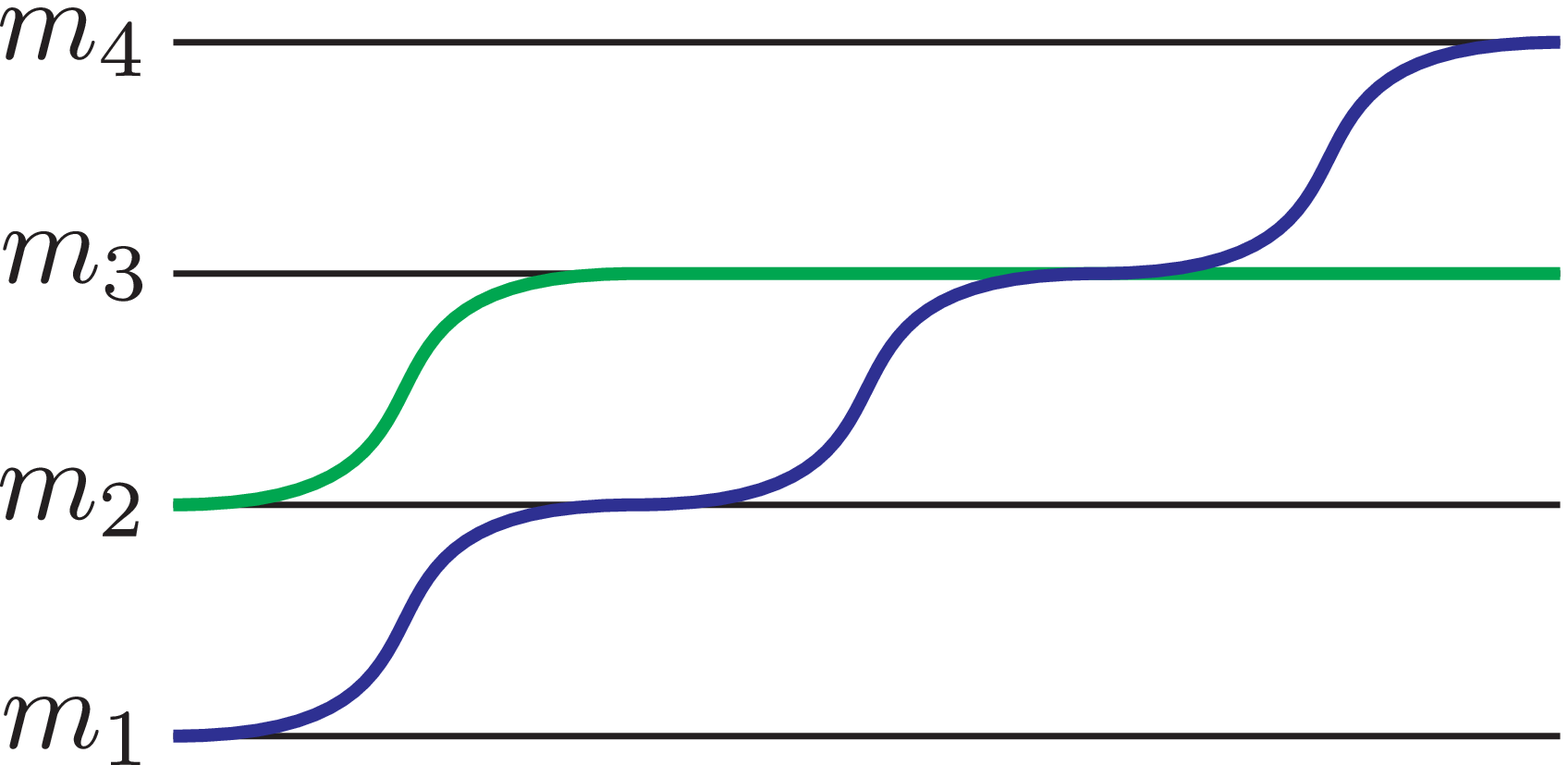}\\
(a) && (b)
\end{tabular}
\end{center}
\caption{The possible domain-wall profiles of the boundary 
condition $\Sigma(-L/2)=\diag(m_1,m_2)$ and 
 $\Sigma(L/2)=\diag(m_3,m_4)$ with the permutations. 
(a) The contribution with no color-line intersections. 
(b) The contribution with an intersection of two color 
lines, which carries a negative sign. 
 }
\label{NAW 12to34}
\end{figure}

Next example is almost the same as the previous case 
except for the boundary condition: 
$\Sigma(-L/2)=\diag(m_1,m_2)$ and 
 $\Sigma(L/2)=\diag(m_3,m_4)$ in the case of $N_c=2$ 
and $N_f=4$. 
Similarly to the previous case, permutations of boundary 
conditions provide two contributions as shown in 
Fig.~\ref{NAW 12to34}.  
Both of them now correspond to BPS configurations 
and are non-vanishing unlike the previous case 
 \bea
 \Volume\left({\mathcal M}^{2,4}_{(1,2) \to (3,4)}\right)
& =&
 \int \frac{d\phi_1}{2\pi}\frac{d\phi_2}{2\pi}
 \frac{ e^{i\beta \phi_1(\hat{L}-(m_3-m_1))}}{(i\phi_1)^3}
\frac{e^{i\beta \phi_2(\hat{L}-(m_4-m_2))}
}{(i\phi_2)^3}\nn\\
&&
 - \int \frac{d\phi_1}{2\pi}\frac{d\phi_2}{2\pi}
 \frac{ e^{i\beta \phi_1(\hat{L}-(m_4-m_1))}}{(i\phi_1)^4}
\frac{e^{i\beta \phi_2(\hat{L}-(m_3-m_2))}
}{(i\phi_2)^2}. 
 \eea
The second term corresponds to the case of two color 
lines intersecting each other, as shown in the 
Fig.~\ref{NAW 12to34}(b). 
Evaluating the $\phi_a$ integral, we obtain
\bea
 \Volume\left({\mathcal M}^{2,4}_{(1,2) \to (3,4)}\right)
&=&\beta^4\Bigg\{
\frac{1}{4}
(\hat{L}-(m_3-m_1))^2
(\hat{L}-(m_4-m_2))^2\nn\\
&&\qquad
-\frac{1}{6}(\hat{L}-(m_4-m_1))^3(\hat{L}-(m_3-m_2))
\Bigg\}.
\eea
This can be expressed by a determinant of the 
$2\times 2$ transition matrix as in 
Eq.(\ref{eq:volume_transition_matrix}) 
\be
{\mathcal T}^{2,4}_{(1,2) \to (3,4)}
=
\begin{pmatrix}
\frac{1}{2!}(\hat{L}-(m_3-m_1))^2 & \frac{1}{3!}(\hat{L}-(m_4-m_1))^3 \\
 \hat{L}-(m_3-m_2) & \frac{1}{2!}(\hat{L}-(m_4-m_2))^2
\end{pmatrix}.
\ee

Let us examine the meaning of our result more closely. 
The kink-profiles such as in 
Fig.~\ref{NAW 12to34} may be understood to 
represent $\Sigma(y)$ connecting vacuum values 
given by boundary conditions. 
Taking for instance the wall connecting $m_3$ and 
$m_4$ in the upper line of Fig.~\ref{NAW 12to34}(a), 
its position can only go to the right up to the other 
wall connecting $m_2$ and $m_3$ in the lower line, 
namely they are non-penetrable each other 
\cite{Isozumi:2004jc,Eto:2006ng}. 
This type of restriction gives an interesting behavior 
of moduli space volume, as illustrated in a concrete 
example in Appendix \ref{app:explicit-comp}. 
On the other hand, our volume formula is given in terms 
of $\phi_a$, the zero mode of $\Phi^a$, which is 
canonically conjugate to the variable $\Sigma^a$. 
The $\phi_a$ integral counts the number of domain walls 
in the $a$-th color line without particular restrictions 
on possible range of wall positions. 
Instead, our formula compensates the over-counting 
of integration range by subtracting appropriate 
contributions in the form of permutations of boundary 
conditions carrying a sign given by the intersection 
number of color lines, as shown in 
Fig.~\ref{NAW 12to34}(b). 
Combining all contributions from permutations of 
boundary conditions, the volume is finally given in 
terms of the determinant of the transition matrix 
(\ref{eq:volume_transition_matrix}).


\section{Duality between Non-Abelian Domain-walls}

We have found a formula for evaluating the volume of the 
domain-wall moduli space.
We here give a non-trivial check of our localization formula 
by examining duality relations between 
two different theories and boundary conditions. 
We take two different gauge theories: 
$G=U(N_c)$ gauge group with $N_f$ flavors and 
$\tilde{G}=U(\tilde{N}_c)$ gauge group with $N_f$ flavors, 
where $\tilde{N}_c\equiv N_f-N_c$. 
The boundary conditions of both theories should be 
chosen to connect complementary vacua as follows. 
If the boundary conditions of the original $G=U(N_c)$ 
theory are $\vec{A}\to\vec{B}$, 
then the corresponding boundary conditions of the other 
$\tilde{G}=U(\tilde{N}_c)$ 
theory should be $\overline{\vec{B}}\to\overline{\vec{A}}$,
where $\overline{\vec{A}}$ and $\overline{\vec{B}}$ are 
the complement of $\vec{A}$ and $\vec{B}$, respectively.
For example, the boundary condition 
$\vec{A}=(2,4,5)$ of $G=U(3)$ theory with $N_f=5$ flavors 
is complementary to the boundary condition 
$\overline{\vec{A}}=(1,3)$ of $\tilde{G}=U(2)$ with $N_f=5$ 
flavors.
Let us call both theories with the complementary 
boundary conditions as dual theories.

In the strong coupling limit, the gauge theories become 
non-linear sigma models and two dual theories become identical 
\cite{Isozumi:2004va, Antoniadis:1996ra}. 
It has been demonstrated explicitly that the moduli spaces of 
domain-walls (in the infinite interval) have a one-to-one 
correspondence and become identical 
in the dual theories in the strong coupling limit 
\cite{Isozumi:2004va}. 
Even in the finite gauge coupling, 
the moduli spaces of 
the domain-walls of these dual theories 
are topologically the same 
\be
{\mathcal M}^{N_c,N_f}_{\vec{A}\to\vec{B}}\simeq
{\mathcal M}^{\tilde{N}_c,N_f}_{\overline{\vec{B}}
\to\overline{\vec{A}}},
\ee
but their metric and other properties are 
different \cite{Isozumi:2004va}. 
Consequently the moduli space of domain-walls in 
these two theories are different at finite gauge coupling, 
but should become identical in the strong coupling limit. 
We need to specify the boundary condition for dual theories. 
The explicit formula of one-to-one correspondence of 
dual theories \cite{Isozumi:2004va} suggests that the 
color-flavor locking of vacua should be chosen in such a way 
that those vacua occupied in dual theories should be the 
complement of each other. 
Namely among $N_f$ flavors, $N_c$ should be 
selected to specify a vacuum in $U(N_c)$ gauge theory, 
whereas the remaining $\tilde{N}_c$ flavors should be selected 
in the dual $U(\tilde{N}_c)$ gauge theory to give the 
dual boundary condition.

We will see that our results for the volume of moduli spaces 
for these two theories differ for finite gauge coupling, 
but become identical for strong coupling limit 
$g^2\to\infty$.

\subsection{Abelian versus non-Abelian duality}

First of all, let us consider duality between Abelian gauge 
group $G=U(1)$ and non-Abelian gauge group 
$\tilde{G}=U(N_f-1)$ with the $N_f$ flavors 
of the same ordered masses $m_1,m_2,\ldots,m_{N_f}$.

For the Abelian model, we take a boundary condition to 
be $\Sigma(-L/2)=m_1$ and $\Sigma(L/2)=m_{N_f}$ to obtain 
the maximal dimensions of the moduli space. 
Using the localization formula of the volume for this
boundary condition, we obtain
\be
\Volume\left({\mathcal M}^{1,N_f}_{1\to N_f}\right)
=\frac{\beta^{N_f-1}}{(N_f-1)!}
\left(\hat{L}-d_{1,N_f}\right)^{N_f-1}.
\label{dual1:Abelian volume}
\ee

The dual of the above Abelian model is the 
$\tilde{G}=U(N_f-1)$ gauge group with $N_f$ flavors. 
The dual boundary condition corresponds also to the maximal 
dimensions of the moduli space and is given by 
$\Sigma^a(-L/2)=m_a$ and  $\Sigma^a(L/2)=m_{a+1}$ 
($a=1,\ldots,N_f-1$). 
The transition matrix of this model becomes 
\begin{multline}
{\mathcal T}^{N_f-1,N_f}_{(1,2,\ldots,N_f-1) \to 
(2,3,\ldots,N_f)}\\
=
\begin{pmatrix}
\hat{L}-d_{12} & \frac{1}{2!}(\hat{L}-d_{13})^2 
& \frac{1}{3!}(\hat{L}-d_{14})^3 & \cdots &
 \frac{1}{(N_f-1)!}(\hat{L}-d_{1,N_f})^{N_f-1}\\
1 & \hat{L}-d_{23} & \frac{1}{2!}(\hat{L}-d_{24})^2 
& \cdots &
 \frac{1}{(N_f-2)!}(\hat{L}-d_{2,N_f})^{N_f-2}\\
0 & 1 & \hat{L}-d_{34} & \cdots &
 \frac{1}{(N_f-3)!}(\hat{L}-d_{3,N_f})^{N_f-3}\\
\vdots & \vdots & \vdots & \ddots & \vdots\\
0 & 0 & 0 & \cdots & \hat{L}-d_{N_f,N_f-1}
\end{pmatrix}.
\label{eq:transition-matrix-Nf-1}
\end{multline}
The volume of the moduli space can be evaluated from the 
determinant of the above transition matrix as 
\be
\Volume\left({\mathcal M}^{N_f-1,N_f}_{(1,2,\ldots,N_f-1) 
\to (2,3,\ldots,N_f)}\right)=
\beta^{N_f-1}\Det {\mathcal T}^{N_f-1,N_f}_{(1,2,\ldots,N_f-1) 
\to (2,3,\ldots,N_f)}.
\label{dual1:non-Abelian volume}
\ee

\begin{figure}
\begin{center}
\begin{tabular}{ccc}
\includegraphics[scale=0.37]{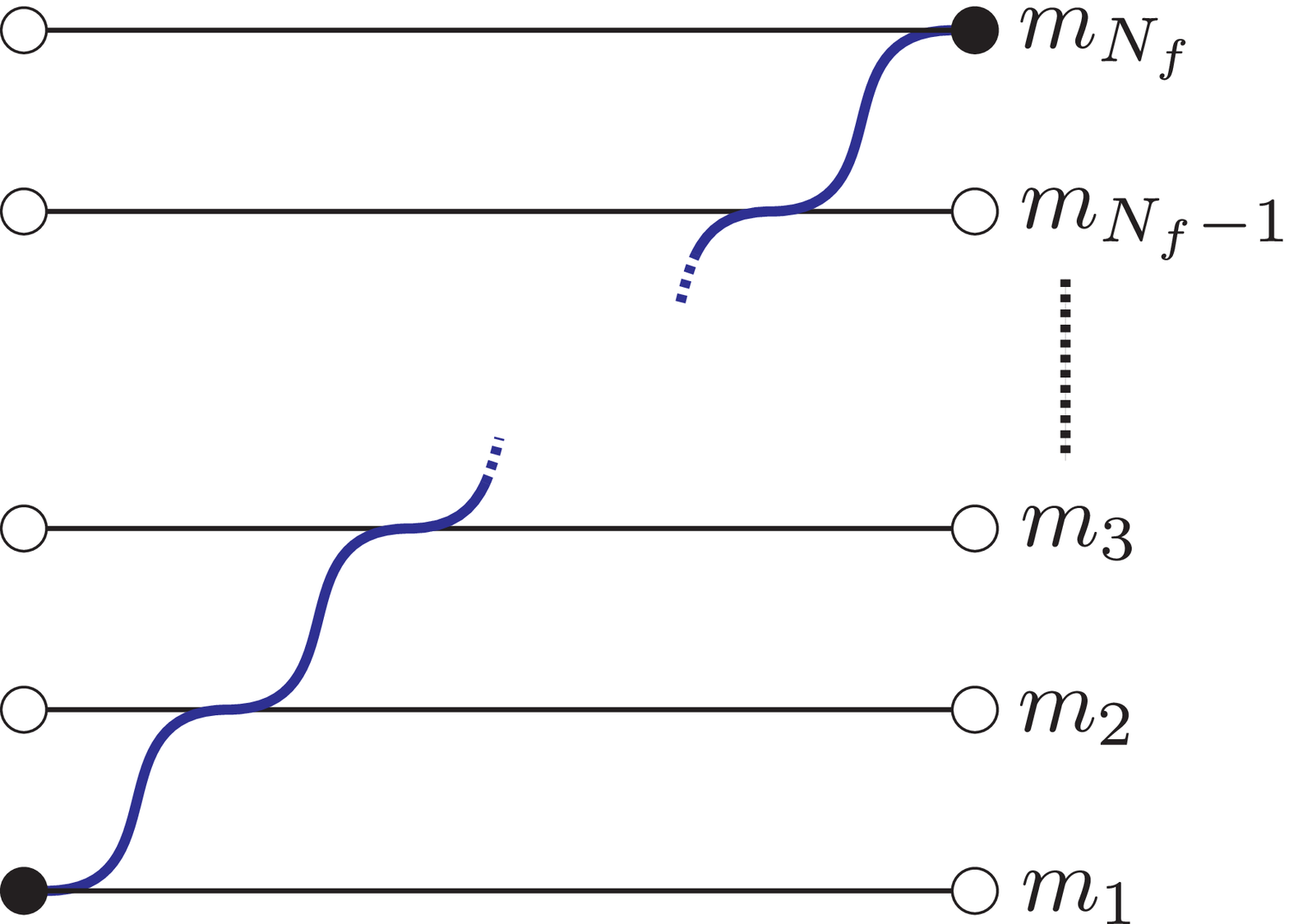} & \hspace{1cm} &
\includegraphics[scale=0.37]{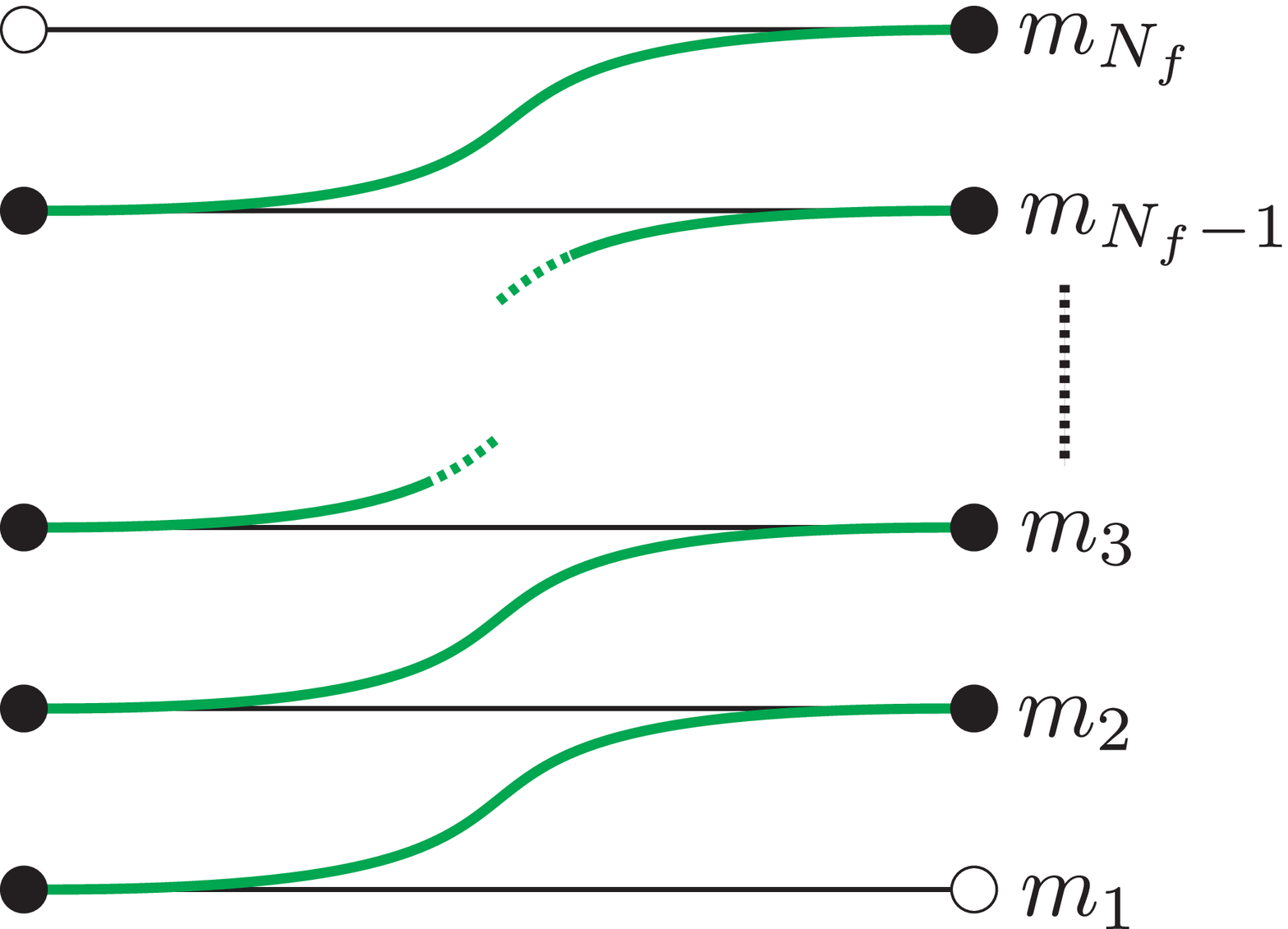}\\
(a) && (b)
\end{tabular}
\end{center}
\caption{The duality between Abelian and non-Abelian theories.
(a) Abelian theory with $N_f$ flavors and the boundary 
condition of $\Sigma(-L/2)=m_1$ and $\Sigma(L/2)=m_{N_f}$.
(b) $G=U(N_f-1)$ non-Abelian theory with $N_f$ flavors
and the boundary condition of 
$\Sigma(-L/2)=\diag(m_1,m_2,m_3,\ldots,m_{N_f-1})$ 
and $\Sigma(L/2)=\diag(m_2,m_3,\ldots,m_{N_f-1},m_{N_f})$.
Black and white circles represent vacua specified by the 
boundary conditions and their complements, respectively.
}
\label{na-dual 1}
\end{figure}

The boundary conditions and typical kink profiles of 
Abelian and non-Abelian theories with $N_f$ flavors 
are depicted in Fig.~\ref{na-dual 1}.

The volume (\ref{dual1:Abelian volume}) and 
(\ref{dual1:non-Abelian volume}) are different from each 
other in a general coupling region. 
In Appendix \ref{app:explicit-comp}, we will explicitly 
demonstrate this difference in the 
simplest case of $N_f=3$ with $N_c=1$ and $N_c=2$ 
as a concrete example. 
We will also show there that the results agree with those of 
a direct calculation using the rigid-rod approximation 
\cite{Eto:2007aw}. 

On the other hand, in the strong coupling limit 
$g^2\to\infty$ where $\hat{L}=\frac{g^2c}{2}L$ becomes 
large, we find 
\be
\Volume\left({\mathcal M}^{1,N_f}_{1\to N_f}\right)
\approx \frac{\beta^{N_f-1}}{(N_f-1)!}\hat{L}^{N_f-1},
\label{vol of CP 1}
\ee
and
\be
\Volume\left({\mathcal M}^{N_f-1,N_f}_{(1,2,\ldots,N_f-1) \to (2,3,\ldots,N_f)}\right)
\approx \frac{\beta^{N_f-1}}{(N_f-1)!}\hat{L}^{N_f-1}.
\label{vol of CP 2}
\ee
(See Appendix \ref{app:nonabelian-duality}.)
Thus the volumes agree with each other in the strong coupling 
limit as expected by the duality.
This result means that the leading terms of the volume in 
$\hat{L}$ coincide in two different models, including a 
combinatorial coefficient. 
This fact is highly non-trivial and suggests our 
localization formula of the volume 
expressed by the determinant works correctly.

In this case, the moduli spaces of both theories are 
topologically isomorphic to 
a complex projective space $\C P^{N_f-1}$
\be
{\mathcal M}^{1,N_f}_{1\to N_f}
\simeq
{\mathcal M}^{N_f-1,N_f}_{(1,2,\ldots,N_f-1) \to 
(2,3,\ldots,N_f)}
\simeq
\C P^{N_f-1}.
\ee
Indeed, the volumes (\ref{vol of CP 1}) and (\ref{vol of CP 2})
represent a rigid volume\footnote{
The power of $\frac{\beta\hat{L}}{2\pi}$ represents a 
{\it complex} dimension of $\C P^{N_f-1}$ 
although it is a real parameter. The volume at the unit 
``size'' is obtained by setting $\frac{\beta\hat{L}}{2\pi}=1$.
} 
of $\C P^{N_f-1}$ with a ``size'' $\frac{\beta\hat{L}}{2\pi}$.

\subsection{Non-Abelian versus non-Abelian duality}

To give another non-trivial check, let us consider a 
duality between two non-Abelian gauge groups.
One model is $G=U(3)$ with $N_f=5$
and the other dual model is $\tilde{G}=U(2)$ with $N_f=5$.
Firstly, we consider
the complementary boundary conditions: 
$\Sigma(-L/2)=\diag(m_1,m_2,m_3)$ and 
$\Sigma(L/2)=\diag(m_3,m_4,m_5)$ for $G=U(3)$ theory,
and 
$\Sigma(-L/2)=\diag(m_1,m_2)$ and $\Sigma(L/2)=\diag(m_4,m_5)$
for $\tilde{G}=U(2)$ theory.
These boundary conditions maximize the dimension of the 
moduli space for the present dual theories.

The transition matrices of both theories are
\be
{\mathcal T}^{3,5}_{(1,2,3) \to (3,4,5)}
=
\begin{pmatrix}
\frac{1}{2!}(\hat{L}-d_{13})^2 & \frac{1}{3!}(\hat{L}-d_{14})^3 & \frac{1}{4!}(\hat{L}-d_{15})^4 \\
\hat{L}-d_{23} & \frac{1}{2!}(\hat{L}-d_{24})^2 & \frac{1}{3!}(\hat{L}-d_{25})^3 \\
1 & \hat{L}-d_{34} & \frac{1}{2!}(\hat{L}-d_{35})^2
\end{pmatrix},
\ee
and
\be
{\mathcal T}^{2,5}_{(1,2) \to (4,5)}
=
\begin{pmatrix}
\frac{1}{3!}(\hat{L}-d_{14})^3 & \frac{1}{4!}(\hat{L}-d_{15})^4 \\
\frac{1}{2!}(\hat{L}-d_{24})^2 & \frac{1}{3!}(\hat{L}-d_{25})^3
\end{pmatrix}.
\ee

\begin{figure}
\begin{center}
\begin{tabular}{ccc}
\includegraphics[scale=0.37]{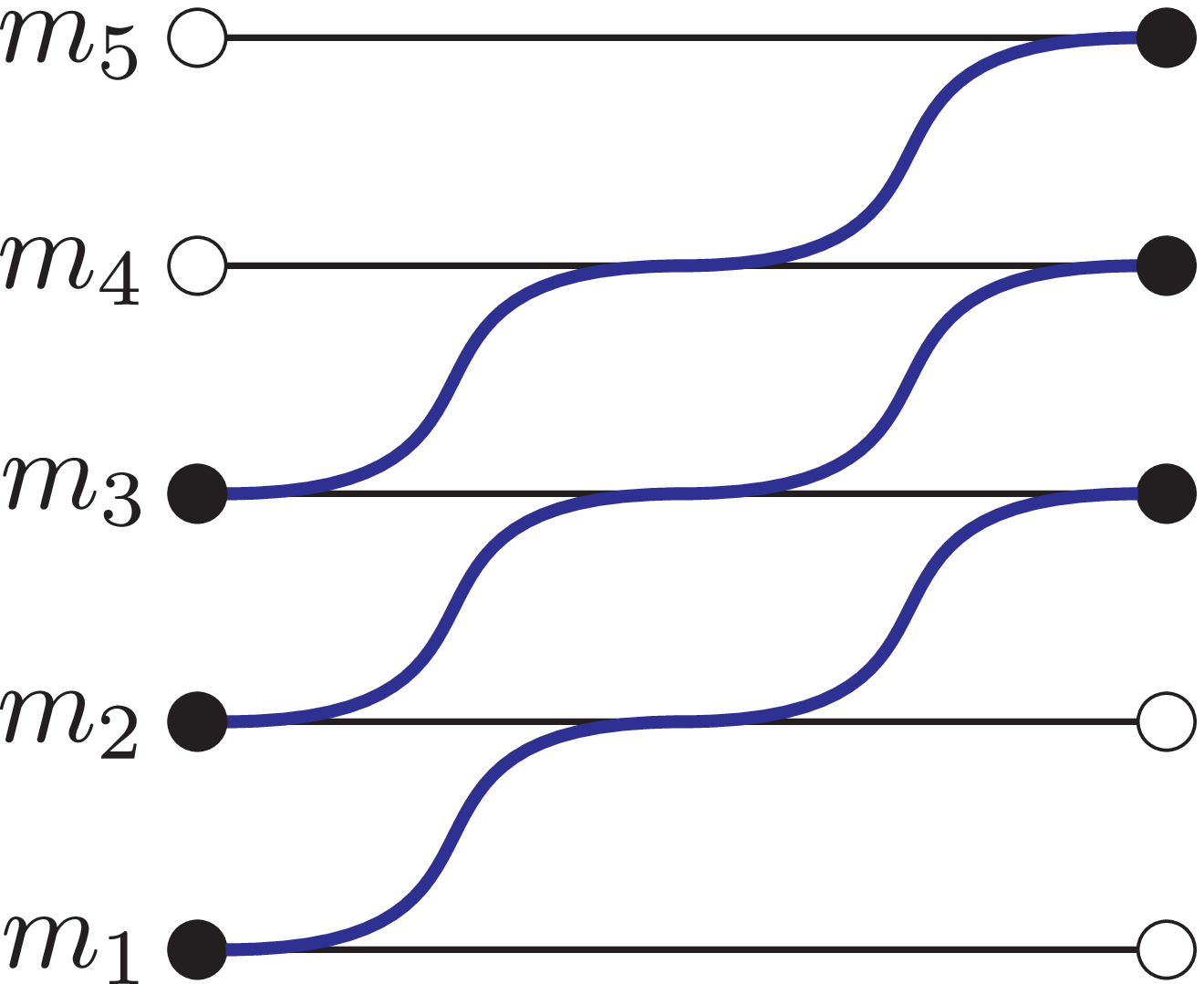} & \hspace{1cm} &
\includegraphics[scale=0.37]{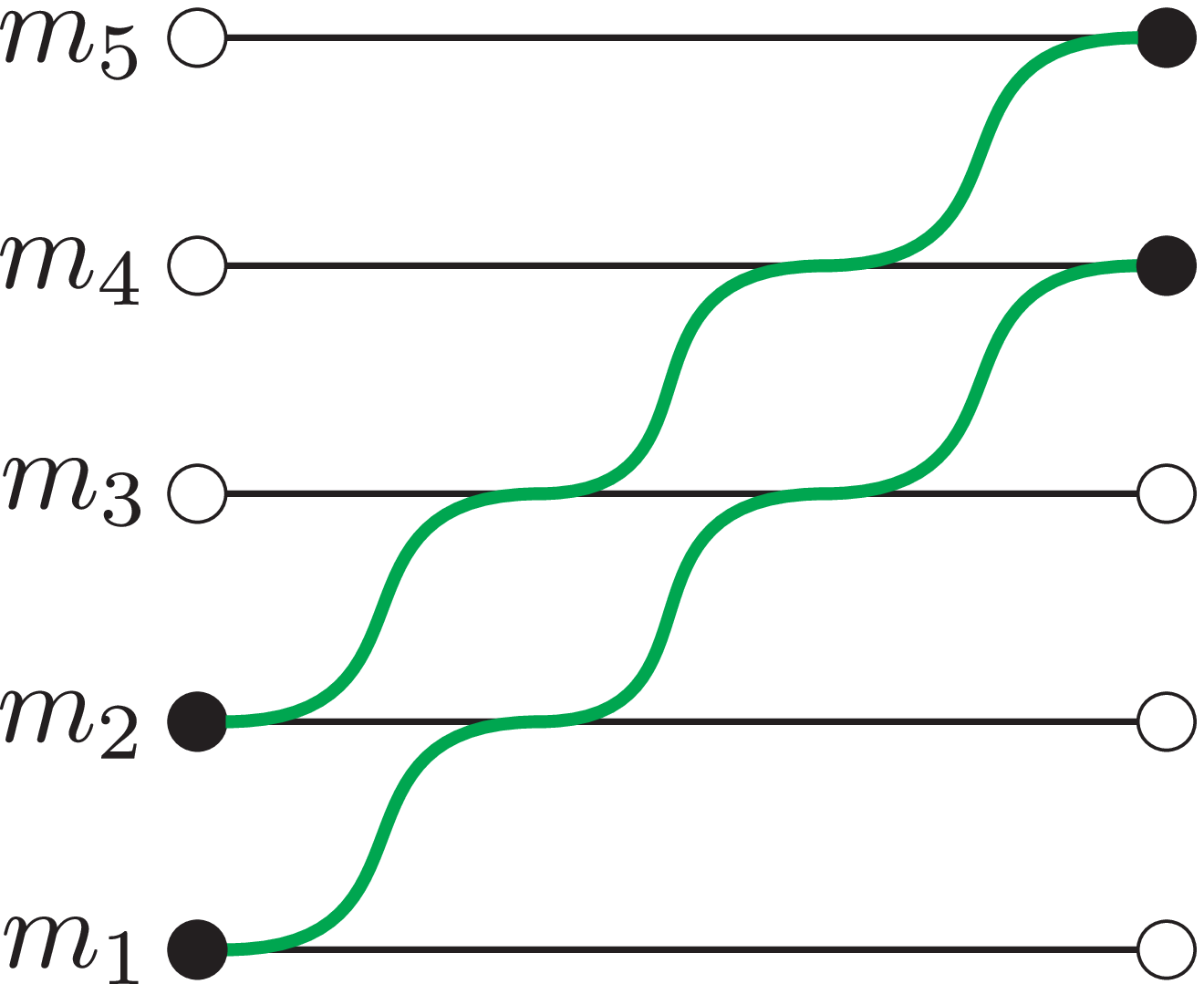}\\
(a) && (b)
\end{tabular}
\end{center}
\caption{The duality between non-Abelian theories.
(a) $G=U(3)$ non-Abelian theory with $N_f=5$ and the boundary condition of $\Sigma(-L/2)=\diag(m_1,m_2,m_3)$ and $\Sigma(L/2)=\diag(m_3,m_4,m_5)$.
(b) $G=U(2)$ non-Abelian theory with $N_f=5$ and the boundary condition of $\Sigma(-L/2)=\diag(m_1,m_2)$
and $\Sigma(L/2)=\diag(m_4,m_5)$.
The boundary conditions are given to connect complementary 
vacua (exchanging the role of black and white circles). 
}
\label{na-dual 2}
\end{figure}

The boundary conditions and typical kink profiles of both 
non-Abelian theories are shown in Fig.~\ref{na-dual 2}.

The volumes of both theories coincide with each other in 
the strong coupling limit
\be
\Volume\left({\mathcal M}^{3,5}_{(1,2,3) \to (3,4,5)}\right)
=\beta^6 \Det {\mathcal T}^{3,5}_{(1,2,3) \to (3,4,5)} = \frac{\beta^6}{144}\hat{L}^6 +{\mathcal O}(\hat{L}^5),
\label{vol of Grassmannian 1}
\ee
and
\be
\Volume\left({\mathcal M}^{2,5}_{(1,2) \to (4,5)}\right)
=\beta^6 \Det {\mathcal T}^{2,5}_{(1,2) \to (4,5)} = \frac{\beta^6}{144}\hat{L}^6 +{\mathcal O}(\hat{L}^5).
\label{vol of Grassmannian 2}
\ee
This result shows that the (complex) dimension\footnote{
Half of the moduli is compact corresponding to 
relative phases of adjacent vacua separated by the 
domain-wall. 
}
of the 
moduli space is 6.

In this maximal dimension case, the moduli spaces are 
isomorphic to a complex Grassmann manifold (Grassmannian)
\be
{\mathcal M}^{3,5}_{(1,2,3) \to (3,4,5)}
\simeq G_{3,5}
\simeq G_{2,5}
\simeq {\mathcal M}^{2,5}_{(1,2) \to (4,5)},
\ee
where $G_{N_c,N_f}$ is expressed by a coset space
\be
G_{N_c,N_f} \equiv \frac{U(N_f)}{U(N_c)\times U(\tilde{N}_c)}.
\ee
The volume of the Grassmannian of unit ``size'' is obtained 
from a quotient of unitary group volumes 
\cite{Macdonald,Fujii,Ooguri:2002gx}
(see also Appendix in \cite{Miyake:2011yr})
\be
\Volume(G_{N_c,N_f})
=\frac{\prod_{j=1}^{N_c}(j-1)! \times \prod_{k=1}^{\tilde{N}_c}(k-1)!}{\prod_{i=1}^{N_f}(i-1)!}
(2\pi)^{N_c\tilde{N}_c}.
\label{eq:grassmann_volume}
\ee 
The volume of the Grassmannian is invariant under 
exchanging $N_c$ and $\tilde{N}_c$.

Using this formula, we notice that the leading term of 
the volumes (\ref{vol of Grassmannian 1}) and 
(\ref{vol of Grassmannian 2}) 
are nothing but the volume of the Grassmannian 
$G_{3,5}$ or $G_{2,5}$ with ``size'' 
$\frac{\beta \hat{L}}{2\pi}$, since 
\be
\Volume(G_{3,5})=\Volume(G_{2,5}) 
=\frac{2!1! \times 1!}{4!3!2!1!}(2\pi)^6
=\frac{1}{144}(2\pi)^6.
\ee
Therefore our results are consistent with the notion that 
the moduli spaces of the 
domain-walls in dual theories asymptotically coincide 
with the Grassmannian $G_{3,5}$ or $G_{2,5}$ with the 
standard metric, but the differential structure (metric) 
is deformed by the sub-leading terms in $\hat{L}$.
These non-trivial agreements strongly suggest that the 
duality between different non-Abelian gauge theories 
is valid in the strong coupling region.

\begin{figure}
\begin{center}
\begin{tabular}{ccc}
\includegraphics[scale=0.37]{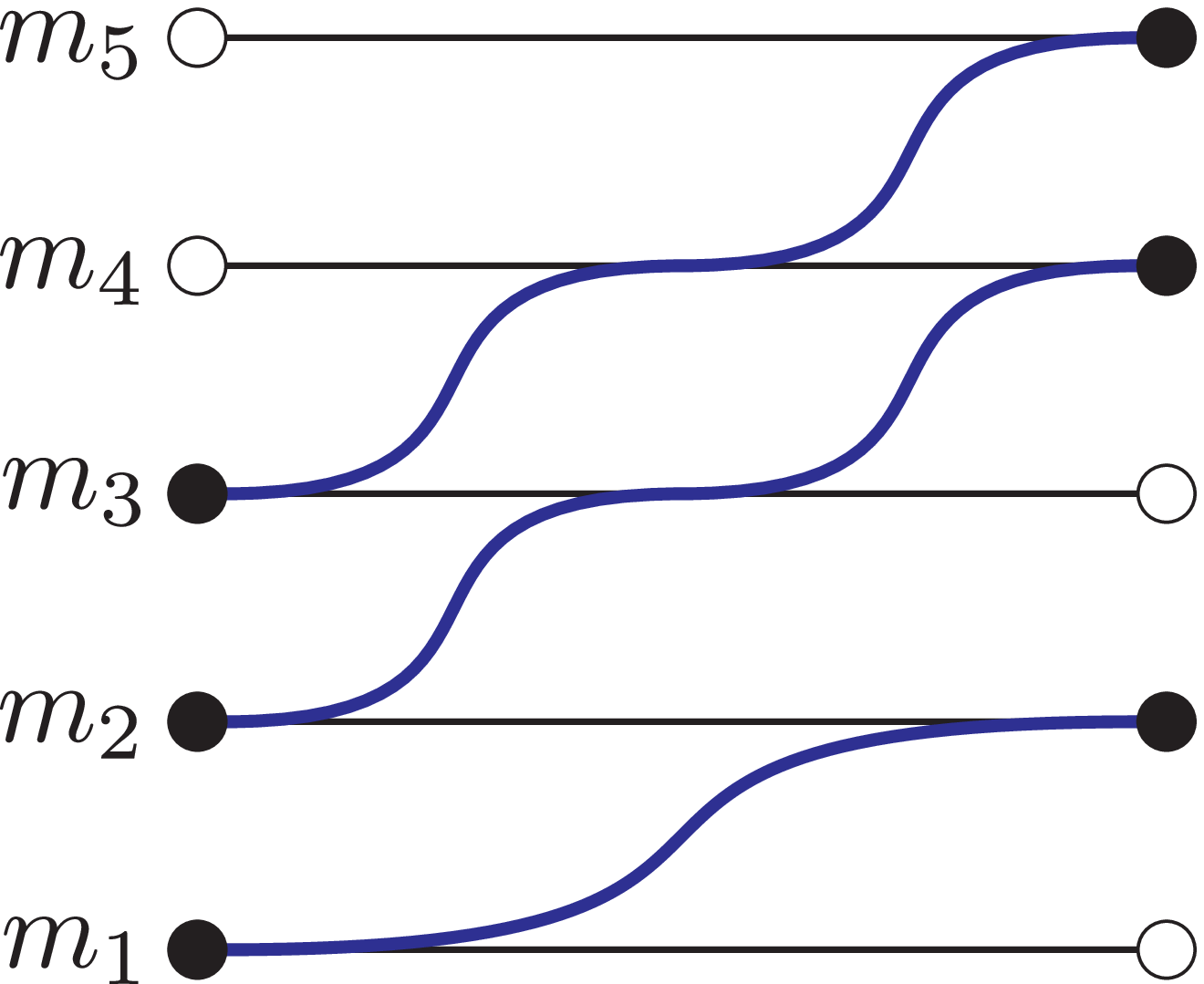} & \hspace{1cm} &
\includegraphics[scale=0.37]{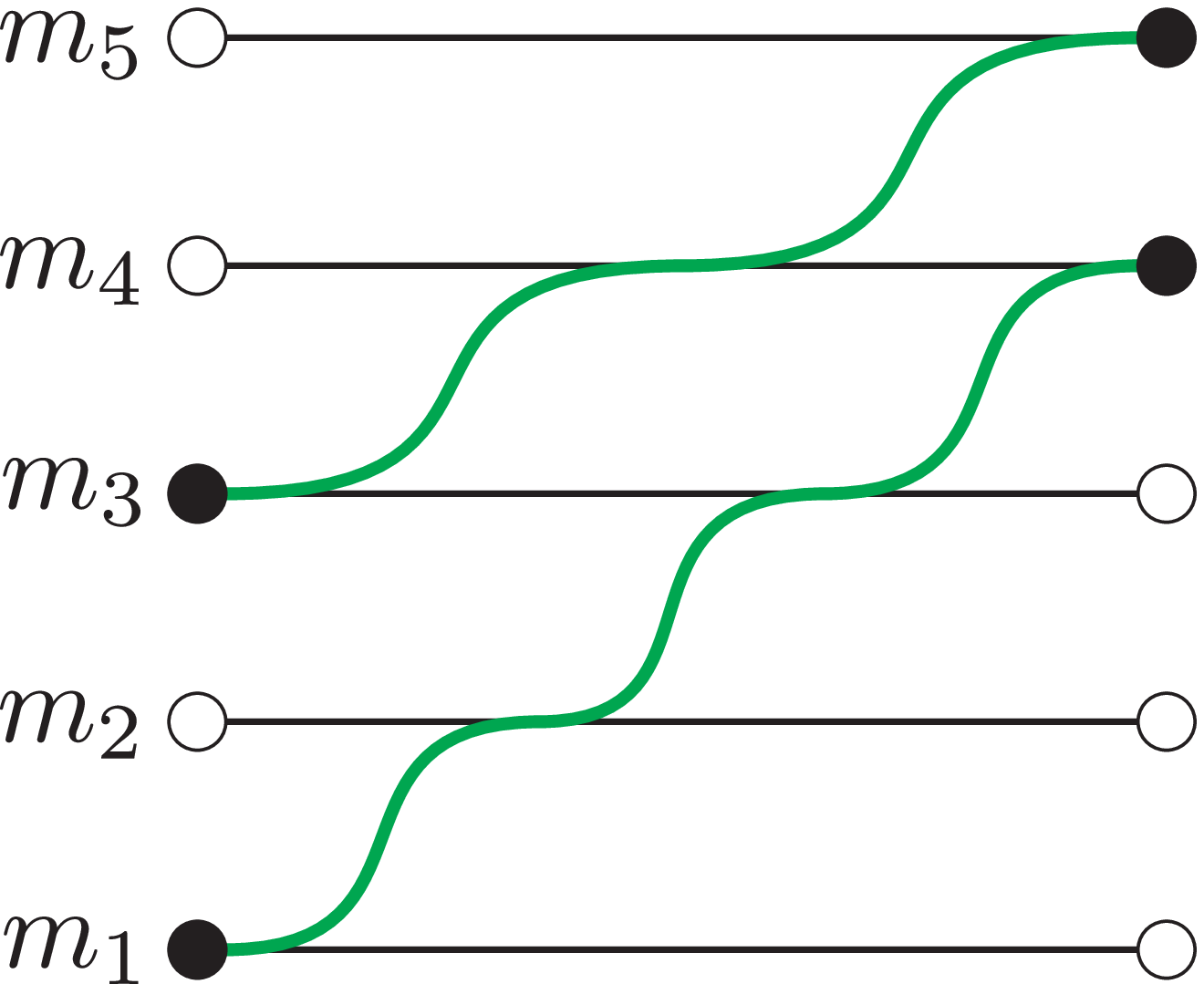}\\
(a) && (b)
\end{tabular}
\end{center}
\caption{The duality between non-Abelian theories with a 
non-maximal boundary condition. 
(a) $G=U(3)$ non-Abelian theory with $N_f=5$ and the boundary condition of $\Sigma(-L/2)=\diag(m_1,m_2,m_3)$ and $\Sigma(L/2)=\diag(m_2,m_4,m_5)$.
(b) $G=U(2)$ non-Abelian theory with $N_f=5$ and the boundary condition of $\Sigma(-L/2)=\diag(m_1,m_3)$
and $\Sigma(L/2)=\diag(m_4,m_5)$.
}
\label{na-dual 3}
\end{figure}

As another example, let us next consider a different 
boundary condition with non-maximal dimensions of 
moduli space : $\Sigma(-L/2)=\diag(m_1,m_2,m_3)$ and 
$\Sigma(L/2)=\diag(m_2,m_4,m_5)$ for $G=U(3)$ theory as the 
boundary condition in one theory, and 
$\Sigma(-L/2)=\diag(m_1,m_3)$ and 
$\Sigma(L/2)=\diag(m_4,m_5)$ for $\tilde{G}=U(2)$ 
theory as the corresponding dual. 
The boundary conditions and typical kink profiles of both 
non-Abelian theories are shown in Fig.~\ref{na-dual 3}.

The transition matrices of both theories with these boundary 
conditions are
\be
{\mathcal T}^{3,5}_{(1,2,3) \to (2,4,5)}
=
\begin{pmatrix}
\hat{L}-d_{12} & \frac{1}{3!}(\hat{L}-d_{14})^3 & \frac{1}{4!}(\hat{L}-d_{15})^4 \\
1 & \frac{1}{2!}(\hat{L}-d_{24})^2 & \frac{1}{3!}(\hat{L}-d_{25})^3 \\
0 & \hat{L}-d_{34} & \frac{1}{2!}(\hat{L}-d_{35})^2
\end{pmatrix},
\ee
and
\be
{\mathcal T}^{2,5}_{(1,3) \to (4,5)}
=
\begin{pmatrix}
\frac{1}{3!}(\hat{L}-d_{14})^3 & \frac{1}{4!}(\hat{L}-d_{15})^4 \\
\hat{L}-d_{34} & \frac{1}{2!}(\hat{L}-d_{35})^2
\end{pmatrix}.
\ee

The volumes of both theories coincide with each other in 
the strong coupling limit 
\be
\Volume\left({\mathcal M}^{3,5}_{(1,2,3) \to (2,4,5)}\right)
=\beta^5 \Det {\mathcal T}^{3,5}_{(1,2,3) \to (2,4,5)} = \frac{\beta^5}{24}\hat{L}^5 +{\mathcal O}(\hat{L}^4),
\ee
and
\be
\Volume\left({\mathcal M}^{2,5}_{(1,3) \to (4,5)}\right)
=\beta^5 \Det {\mathcal T}^{2,5}_{(1,3) \to (4,5)} 
= \frac{\beta^5}{24}\hat{L}^5 +{\mathcal O}(\hat{L}^4).
\ee
This result shows that the (complex) dimension of the moduli 
space is 5, which is smaller than the maximal dimension 6, 
as expected.
So the moduli space for the present boundary conditions 
should be a complex sub-manifold 
of the Grassmannian $G_{3,5}\simeq G_{2,5}$.

In Appendix \ref{app:nonabelian-duality}, we evaluate 
the asymptotic behavior of the volume of 
the moduli space in the case of maximal dimensions for the 
general $N_c$ gauge theories with $N_f$ flavors to 
 obtain 
\be
\begin{split}
\Volume\left({\mathcal M}^{N_c,N_f}_{(1,2,\ldots,N_c)\to(\tilde{N}_c+1,\ldots,N_f-1,N_f)}\right)
&=\frac{\prod_{j=1}^{N_c}(j-1)! \times \prod_{k=1}^{\tilde{N}_c}(k-1)!}{\prod_{i=1}^{N_f}(i-1)!}
\left(\beta\hat{L}\right)^{N_c\tilde{N}_c}+\cdots,\\
\Volume\left({\mathcal M}^{\tilde{N}_c,N_f}_{(1,2,\ldots,\tilde{N}_c)\to(N_c+1,\ldots,N_f-1,N_f)}\right)
&=\frac{\prod_{j=1}^{N_c}(j-1)! \times \prod_{k=1}^{\tilde{N}_c}(k-1)!}{\prod_{i=1}^{N_f}(i-1)!}
\left(\beta\hat{L}\right)^{N_c\tilde{N}_c}+\cdots.
\end{split}
\label{eq:leading_duality}
\ee
This result shows 
that there exists a duality relation between two 
different domain-wall theories in the strong coupling region.

\section{T-duality to Vortex on Cylinder}

In this section, we discuss another kind of duality between 
the domain-walls and vortices. 
As discussed in \cite{Eto:2006mz,Eto:2007aw}, there exists 
a T-duality relation between vortices on a cylinder 
and domain-walls on the interval.
We here would like to show that the volume of the moduli space 
exhibits this T-duality. 
As a base space we consider a cylinder, which is 
a two-dimensional surface of a circle $S^1$ with the 
radius $\beta$ times an interval $I$ with the length $L$.

\begin{figure}
\begin{center}
\begin{tabular}{ccc}
\includegraphics[scale=0.38]{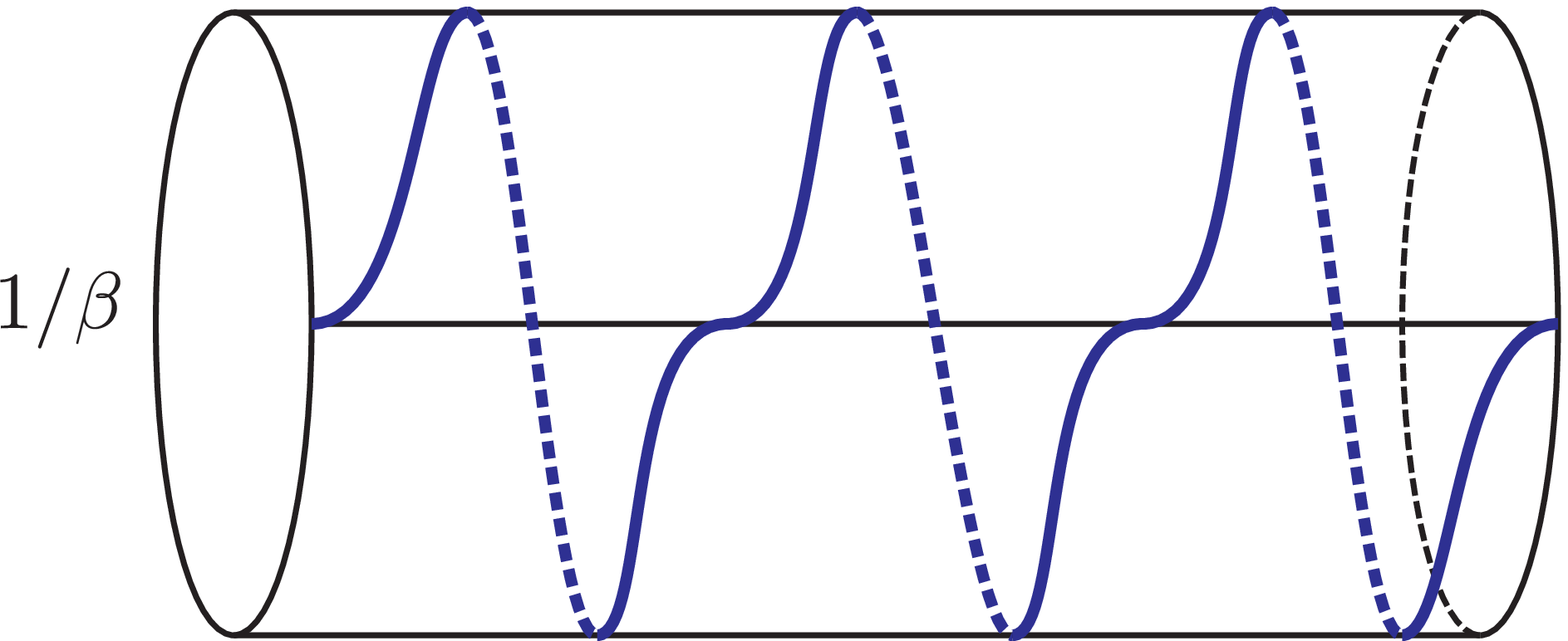} & \raisebox{1.4cm}{$\Rightarrow$} &
\includegraphics[scale=0.38]{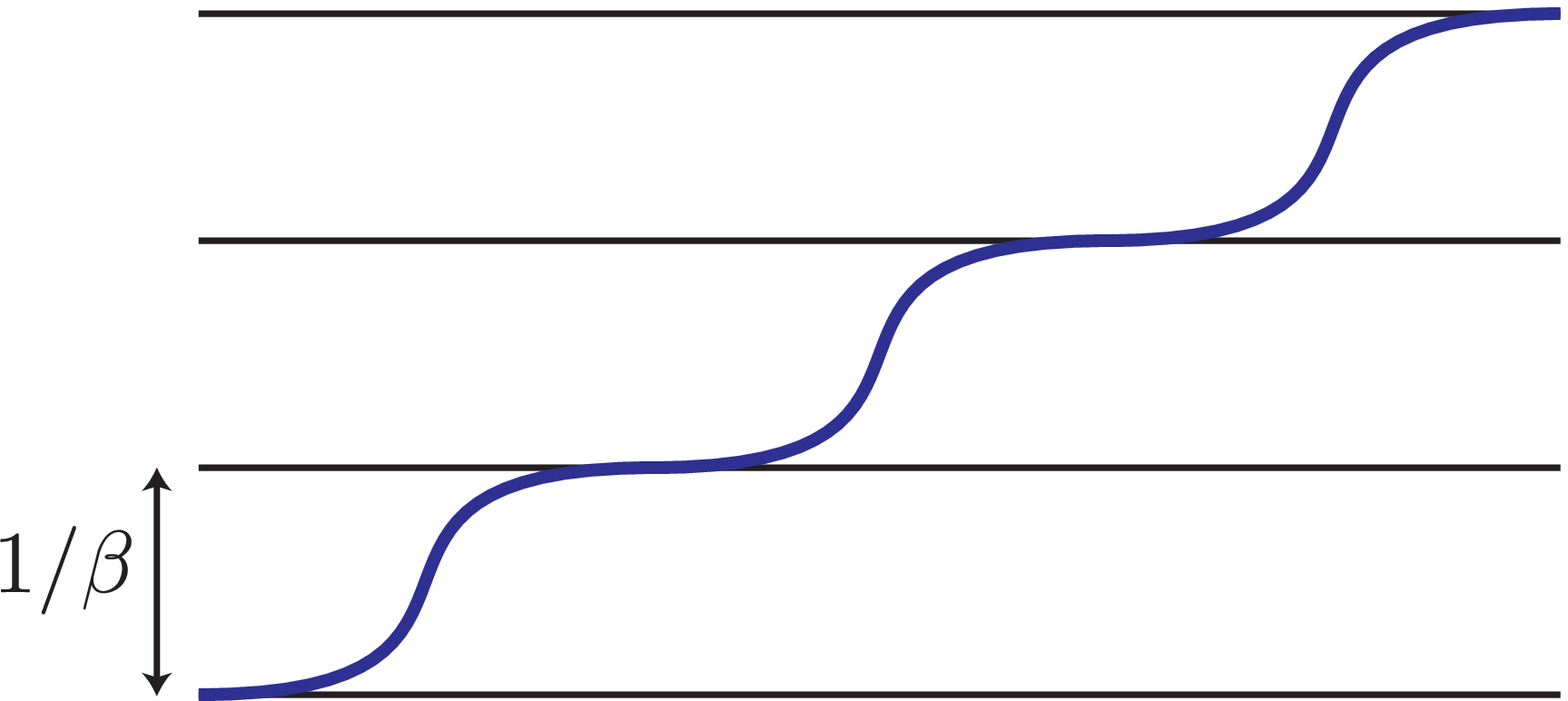}\\
(a) && (b)
\end{tabular}
\end{center}
\caption{T-dual picture of the vortex on a cylinder. 
The $k$ vortices on the cylinder are dual to the 
domain-walls wrapping $k$-times around the circle. 
(a) A covering space of the $k$ domain-walls on the 
cylinder. (b) The $k$ domain-walls in 
the infinite number of flavors of the mass difference 
$1/\beta$. They are equivalent. 
}
\label{vortex dual}
\end{figure}

To see this duality, we start with the simplest case: 
vortices in $U(1)$ gauge theory with a single charged 
matter, which are called Abelian local vortices, 
or Abrikosov-Nielsen-Olesen (ANO) vortices \cite{Abrikosov:1956sx}. 
If there are $k$ vortices on the cylinder, 
the vortices are mapped to $k$ domain-walls (kinks) on 
the interval with the length $L$ by the T-duality.
The charged matters are mapped to the matter branes 
\cite{Eto:2004vy} and we can 
regard mass differences for each kink to be 
$1/\beta$, which is the radius of the dual circle in the 
domain-wall picture. (See Fig.~\ref{vortex dual}.)

The total mass difference between the boundary conditions 
at $y=-L/2$ and at $y=L/2$ is $k/\beta$. 
So we can derive the integral formula for the volume of 
this domain-wall moduli space as 
\bea
 \Volume\left({\mathcal M}^{1,1}_k\right)
 &=&\int_{-\infty}^\infty \frac{d\phi}{2\pi}
 \frac{1}{(i\phi)^{k+1}}e^{i\phi \beta 
\left(\hat{L}-\frac{k}{\beta}\right)}\nn\\
 &=& \frac{1}{k!} \left(\beta \hat{L}-k\right)^k.
 \label{abelian vortex}
\eea
Recalling that the area of the cylinder in the vortex picture 
is given by ${\mathcal A}= 2\pi \beta L$ and that 
$\hat{L}=g^2cL/2$ in Eq.(\ref{eq:def-hat-L}), 
the volume (\ref{abelian vortex}) is equivalent to the 
volume of the ANO vortices on the cylinder
\bea
 \Volume\left({\mathcal M}^{1,1}_k \right)
 = \frac{1}{k!} \left(\frac{g^2 c}{4\pi}
{\mathcal A}-k\right)^k,
\eea
where we can regard $g$ and $c$ as the gauge coupling 
and the FI parameter in the two-dimensional vortex system, 
respectively, since the combination $g^2c$ is invariant 
under the T-duality.
In the large area limit ${\mathcal A}\to \infty$, the 
volume is proportional to ${\mathcal A}^k/k!$, 
which is the volume of the symmetric product space of the 
cylinder $(S^1\times I)^k/\mathfrak{S}_k$. 
This is consistent with the point-like behavior of the vortex 
in the large area limit.

Now let us consider the Abelian $k$ vortices with $N_f$ 
matter fields of identical charges. 
This is called Abelian semi-local vortices. 
In the vortex side, the masses of the charged matters are 
degenerate and they are T-dual to degenerate vacua 
in the domain-wall picture.
Since it is subtle to treat the degenerate masses in 
the domain-wall side \cite{Eto:2008dm}, 
we split the masses of the $N_f$ matters by giving small 
mass differences $\varepsilon$.

\begin{figure}
\begin{center}
\includegraphics[scale=0.5]{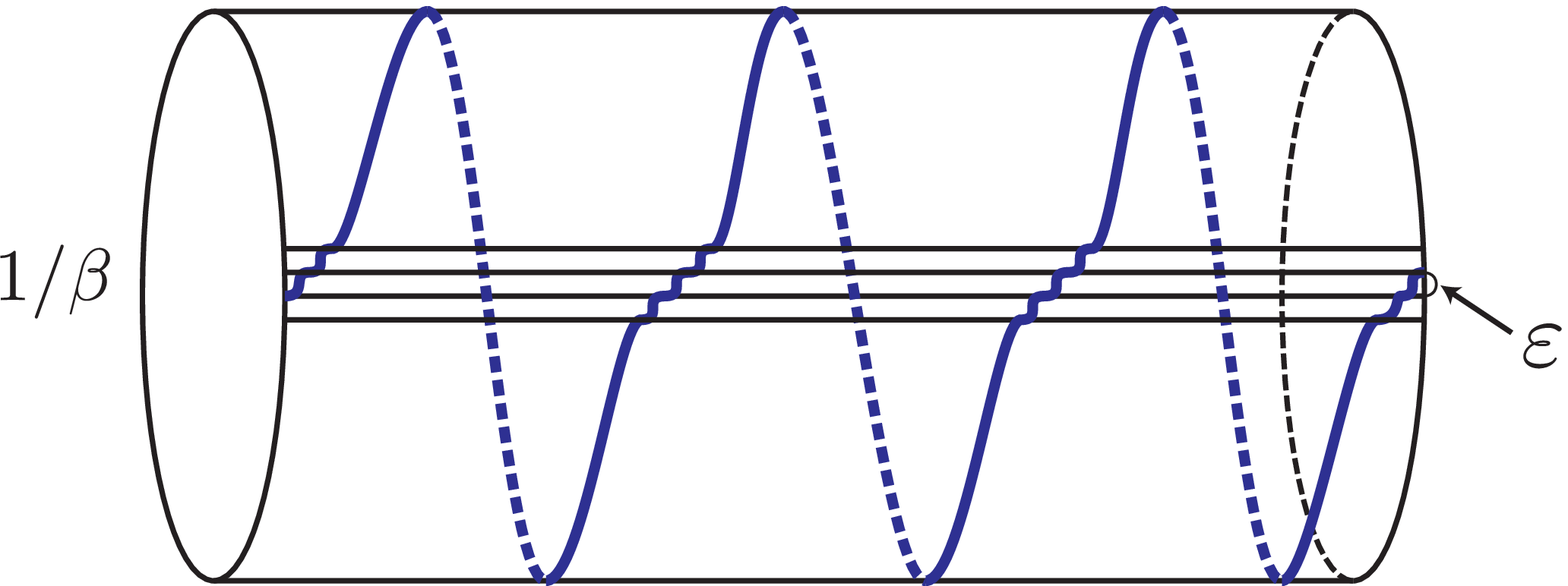} \\
\vspace{1cm}
\includegraphics[scale=0.6]{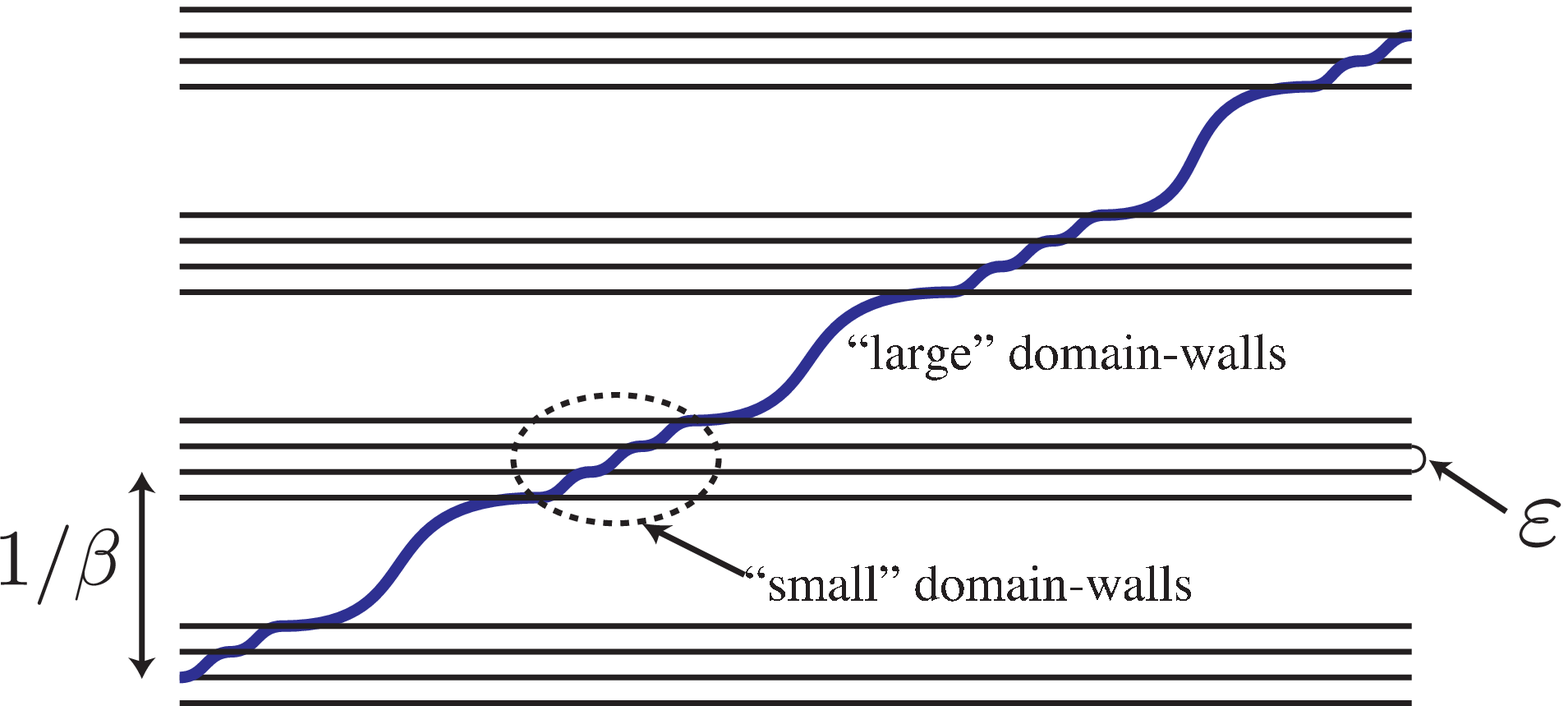}
\end{center}
\caption{T-dual picture of the vortex on the cylinder with 
$N_f>1$ (Abelian semi-local vortex). 
The $k$ vortices on the cylinder are dual to the large
domain-walls wrapping $k$-times around the circle,
and the multiple $N_f$ flavors produce the small domain-walls
with the small mass differences $\varepsilon$.
We depict the case of $N_f=4$ and $k=3$.
The number of the small domain-walls is
10 and the total number of the domain-walls is 
$10+3 = 3\times 4 +1$, that is, $n=1$ in this example.
}
\label{vortex dual with flavor}
\end{figure}

There are two different types of domain-walls in this $N_f$ 
flavor case.
One type comes from the $k$ vortices, which becomes ``large'' 
domain-walls with the mass difference $1/\beta$.
The other type is ``small'' domain-walls connecting the 
small mass differences $\varepsilon$.
The number of the large domain-walls is always $k$, since 
they are $k$ winding domain-walls around the cylinder.
The number of small domain-walls varies from 
$(k-1)\times (N_f-1)$ to $(k+1)\times (N_f-1)$,
depending on the boundary conditions at $y=-L/2$ and $y=L/2$.
So the total number of domain-walls with $N_f$ charged matters 
varies from $k N_f -(N_f-1)$ to $k N_f +(N_f-1)$, where 
we have assumed $k>0$. 
We denote this number by $k N_f+n$, where 
$n=-(N_f-1),\ldots, +(N_f-1)$ if $k>0$
\footnote{When $k=0$, $n$ runs from 0 to $+(N_f-1)$}.
This means that the index for the domain-walls is $kN_f+n+1$.
Note here that $n$ is determined by the number of the small 
domain-walls adjacent to the boundaries 
(boundary condition of the domain-walls).
An example of the domain-wall configuration is shown in 
Fig.~\ref{vortex dual with flavor}.

Noting that the mass difference of each one of the large 
and small domain-wall is $1/\beta-(N_f-1)\varepsilon$ and 
$\varepsilon$, respectively, we find the total mass 
difference of $k$ large domain-walls and $k(N_f-1)+n$ small 
domain-walls is 
$k\times(1/\beta-(N_f-1)\varepsilon)+(k(N_f-1)+n)\times 
\varepsilon=k/\beta+n \varepsilon$. 
Then, applying the localization formula to the above 
domain-wall configuration,
we obtain the volume formula
\bea
 \Volume\left({\mathcal M}^{1,N_f}_k (S^1\times I)\right)
 &=&\int_{-\infty}^\infty \frac{d\phi}{2\pi}
 \frac{1}{(i\phi)^{k N_f+n+1}}e^{i\phi \beta \left(\hat{L}-\frac{k}{\beta}
 -n\varepsilon\right)}\nn\\
 &=& \frac{1}{(kN_f+n)!} \left(\beta \hat{L}-k
  -n\beta\varepsilon \right)^{kN_f+n}\nn\\
 &=& \frac{1}{(kN_f+n)!} \left(\A-k
  -n\beta\varepsilon \right)^{kN_f+n},
\eea
where we have defined $\A\equiv\beta \hat{L} = \frac{g^2c}{4\pi}{\mathcal A}$.
In the $\varepsilon\to0$ limit, we find
\be
 \Volume\left({\mathcal M}^{1,N_f}_k(S^1\times I)\right)
= \frac{1}{(kN_f+n)!} \left(\A-k
  \right)^{kN_f+n}.
\ee
We can see that the above volume is the same as the volume 
of the moduli space of Abelian semi-local vortices with 
$N_f$ flavors on the sphere \cite{Miyake:2011yr} if $n=N_f-1$.

In the large area limit ${\mathcal A}\to\infty$, the volume 
of the moduli space of the vortices on the cylinder 
(dual to the large and small domain-walls) is 
proportional to ${\mathcal A}^{kN_f+n}$.
We do not know an explicit formula for the volume of the 
moduli space of the vortices on the cylinder,
but this large area behavior suggests that the dimension of 
the moduli space of the vortex is $N_f+n$ and the index of 
the operator ${\mathcal D}_\zb$ on the cylinder with the 
appropriate boundary condition, which counts the number of 
zero modes of the Higgs fields obeying 
${\mathcal D}_\zb H=0$ and determines the power of 
${\mathcal A}$ via the contour integral, is $N_f+n+1$.
So we expect that the index of the operator 
${\mathcal D}_\zb$ on the cylinder may be given by 
the Atiyah-Patodi-Singer index 
theorem \cite{AtiyahPatodiSinger}
\be
\begin{split}
\ind {\mathcal D}_\zb &= N_f \int_{S^1\times I}F 
-\frac{N_f}{2}\left[\eta(S^1_R)-\eta(S^1_L)\right]\\
&=kN_f+\left\lfloor N_f \left(\oint_{S^1_R} A
-\oint_{S^1_L} A\right) \right\rfloor + 1\\
&=kN_f + n +1,
\end{split}
\label{eq:index-vortex}
\ee
where  $S^1_R$ and $S^1_L$ are the right and left boundaries of 
the cylinder, respectively, $\eta$ is the eta-invariant at 
the boundaries, and $\lfloor x \rfloor$ stands for the floor 
function which gives the largest integer not greater than $x$.
The index theorem implies that the value of $n$ in 
Eq.(\ref{eq:index-vortex}) for vortices on the cylinder is 
also limited to be $-(N_f-1) \leq n \leq +(N_f-1)$ because 
of the T-duality. 
We expect that $n$ is determined by the holonomies at 
the boundaries of the cylinder. To see a precise 
correspondence between $n$ and holonomies, 
we need further investigation of the moduli of the vortex 
on the cylinder.

Finally, we discuss an extension of the above observations 
in the Abelian case to the non-Abelian case.
As explained in the previous sections, the evaluation of 
the volume of the non-Abelian domain-wall moduli space 
can be reduced to a sum of products of the Abelian ones. 
So the T-dualized domain-walls of the non-Abelian vortex 
can also be decomposed into the Abelian ones.
In this decomposition, we have to take into account the 
permutations of the boundary conditions for each 
Abelian component. The boundary condition is labeled by 
the integer $n$, which reflects the different number 
of the small domain-walls, as explained above. 
Thus each Abelian part of the domain-walls is labeled by 
the decomposed vortex charge $k_a$ and the integer $n_a$ 
associated with the boundary conditions, where $a$ runs over 
the rank of $U(N_c)$, namely $a=1,\ldots, N_c$ and $k_a$ 
satisfies $k=\sum_{a=1}^{N_c}k_a$.

Thus, using the decomposition, we obtain the localization formula for the volume of the moduli space
of the T-dualized non-Abelian vortex on the cylinder
\bea
 \Volume\left({\mathcal M}^{N_c,N_f}_k(S^1\times I)\right)
 &=&  \sum_{ \vec{k},\vec{n}}
  (-1)^{|\sigma(\vec{k},\vec{n})|}
 \prod_{a=1}^{N_c}
 \int_{-\infty}^\infty \frac{d\phi_a}{2\pi}
 \frac{1}{(i\phi_a)^{k_a N_f+n_a +1}}e^{i\phi_a \beta \left(\hat{L}-\frac{k_a}{\beta}-n_a \varepsilon\right)}\nn\\
 &=&  \sum_{ \vec{k},\vec{n}}
  (-1)^{|\sigma(\vec{k},\vec{n})|}
  \prod_{a=1}^{N_c}\frac{1}{(k_a N_f+n_a)!} \left(\beta \hat{L}-k_a-n_a \beta \varepsilon\right)^{k_aN_f+n_a}\nn\\
 &=&  \sum_{ \vec{k},\vec{n}}
  (-1)^{|\sigma(\vec{k},\vec{n})|}
  \prod_{a=1}^{N_c}\frac{1}{(k_a N_f+n_a)!} \left(\A-k_a-n_a \beta \varepsilon\right)^{k_aN_f+n_a},\nn\\
\eea
where $\vec{k}$ are all possible $N_c$ component integer 
vectors satisfying $k=\sum_{a=1}^{N_c}k_a$ and with an 
ordering of $k_1\geq k_2 \geq \cdots \geq k_{N_c}$, and 
$\vec{n}$ are varied in the given boundary conditions. 
The signature of each term depends on the order of the 
permutations $\sigma(\vec{k},\vec{n})$ of the boundary 
conditions determined by $\vec{k}$ and $\vec{n}$. 
The signature $(-1)^{|\sigma(\vec{k},\vec{n})|}$ 
is given by the parity of 
the intersection number of the $N_c$ color lines. 
The volume of the moduli space of the non-Abelian vortex 
on the cylinder may be obtained in the limit of
$\varepsilon \to0$. 
This formula also should be directly checked from the 
localization theorem for the vortex on 
the cylinder with the boundary conditions of the various 
holonomies.

To see the above construction concretely, let us consider 
only an example of $N_c=2$ and general $N_f$ for simplicity 
in the following, since the number of charge partitions 
increases rapidly for large $N_c$. 
We also take a trivial boundary condition, namely, 
$(1,2)\to(1,2)$. 

First, for $k=0$, there is no partition of the charges, 
namely $(k_1,k_2)=(0,0)$.
And also there is no choice of the boundary conditions. 
In the T-dual picture of domain-walls, 
this means that there exists no domain-wall, but
 the volume of the moduli space gives a finite contribution 
\be
 \Volume\left({\mathcal M}^{2,N_f}_0(S^1\times I)\right)
 = 1.
\label{eq:vacuum-sector}
\ee
This should provide the relative normalization of the volume.

For $k=1$, there are two partitions of the charges, which are $\vec{k}=(1,0)$.
For this partition, there are two permutations of the boundary conditions, that give
$\vec{n}=\{(-1,+1),(0,0)\}$. 
Thus the summation over all possible combinations of the charges and the boundary conditions
gives the volume of the moduli space of the non-Abelian domain-walls
\be
\begin{split}
 \Volume\left({\mathcal M}^{2,N_f}_1(S^1\times I)\right)
 &=  -\frac{2}{N_f!} \left(\hat{\mathcal A}-1\right)^{N_f}
 +\frac{1}{(N_f-1)!}\left(\hat{\mathcal A}-1+\hat\varepsilon\right)^{N_f-1} \left(\hat{\mathcal A}-\hat\varepsilon\right)
\end{split}
\ee
if $\hat{\mathcal A}>1$, where we define 
$\hat\varepsilon \equiv \beta \varepsilon$. The volume of the moduli space of the vortices
is obtained in the limit of $\hat\varepsilon\to 0$.

For $k=2$, we have the choices of the charges and boundary conditions as
$\vec{k}=\{(2,0),(1,1)\}$ and $\vec{n}=\{(0,0),(-1,1)\}$.
Then the volume becomes
\be
\begin{split}
 \Volume\left({\mathcal M}^{2,N_f}_2(S^1\times I)\right)
 &= \frac{2}{(2N_f)!} \left(\hat{\mathcal A}-2\right)^{2N_f}\\
 &\quad-\frac{1}{(2N_f-1)!}\left(\hat{\mathcal A}-2+\hat\varepsilon\right)^{2N_f-1}\left(\hat{\mathcal A}-\hat\varepsilon\right)\\
 &\quad+\frac{1}{N_f!N_f!}\left(\hat{\mathcal A}-1\right)^{2N_f}\\
 &\quad
 -\frac{1}{(N_f+1)!(N_f-1)!}\left(\hat{\mathcal A}-1-\hat\varepsilon\right)^{N_f+1}\left(\hat{\mathcal A}-1+\hat\varepsilon\right)^{N_f-1},
\end{split}
\ee
if $\hat{\mathcal A}>2$.

Similarly, we obtain
\be
\begin{split}
 \Volume\left({\mathcal M}^{2,N_f}_3(S^1\times I)\right)
&=-\frac{2}{(3N_f)!}\left(\A-3\right)^{3N_f}\\
&\quad+\frac{1}{(3N_f-1)!}\left(\A-3+\hat\varepsilon\right)^{3N_f-1}\left(\A-\hat\varepsilon\right)\\
&\quad-\frac{2}{(2N_f)!N_f!}\left(\A-2\right)^{2N_f}\left(\A-1\right)^{N_f}\\
&\quad+\frac{1}{(2N_f+1)!(N_f-1)!}\left(\A-2-\hat\varepsilon\right)^{2N_f+1}\left(\A-1+\hat\varepsilon\right)^{N_f-1}\\
&\quad+\frac{1}{(2N_f-1)!(N_f+1)!}\left(\A-2+\hat\varepsilon\right)^{2N_f-1}\left(\A-1-\hat\varepsilon\right)^{N_f+1}
\end{split}
\ee
if $\hat{\mathcal A}>3$ for $k=3$, and
\be
\begin{split}
 \Volume\left({\mathcal M}^{2,N_f}_4(S^1\times I)\right)
 &=\frac{2}{(4N_f)!}\left(\A-4\right)^{4N_f}\\
&\quad-\frac{1}{(4N_f-1)!}\left(\A-4+\hat\varepsilon\right)^{4N_f-1}\left(\A-\hat\varepsilon\right)\\
&\quad+\frac{2}{(3N_f)!N_f!}\left(\A-3\right)^{3N_f}\left(\A-1\right)^{N_f}\\
&\quad-\frac{1}{(3N_f+1)!(N_f-1)!}\left(\A-3-\hat\varepsilon\right)^{3N_f+1}\left(\A-1+\hat\varepsilon\right)^{N_f-1}\\
&\quad-\frac{1}{(3N_f-1)!(N_f+1)!}\left(\A-3+\hat\varepsilon\right)^{3N_f-1}\left(\A-1-\hat\varepsilon\right)^{N_f+1}\\
&\quad+\frac{1}{(2N_f)!(2N_f)!}\left(\A-2\right)^{4N_f}\\
&\quad-\frac{1}{(2N_f+1)!(2N_f-1)!}\left(\A-2-\hat\varepsilon\right)^{2N_f+1}\left(\A-2+\hat\varepsilon\right)^{2N_f-1},
\end{split}
\label{k=4 A>4}
\ee
if $\hat{\mathcal A}>4$ for $k=4$.

So far, we have considered the general number of flavors 
$N_f$ with the $U(2)$ gauge group.
The volume of the moduli space of $k$ vortices is a 
complicated $kN_f$-th order polynomial in $\A$.
However, setting $N_f=N_c=2$, we see the order of polynomial 
of the volume remarkably reduces. 
The vortices in this situation ($N_f=N_c$) is called 
non-Abelian local vortices. 

Putting $N_f=N_c=2$ into the results 
(\ref{eq:vacuum-sector})-(\ref{k=4 A>4}) 
for general $N_f$, 
we find the volume of the moduli space of non-Abelian local 
vortices on the cylinder as 
\bea
\Volume\left({\mathcal M}^{2,2}_0(S^1\times I)\right)
 &=& 1,\nn\\
\Volume\left({\mathcal M}^{2,2}_1(S^1\times I)\right)
 &=&\hat{\mathcal A} -1+\hat\varepsilon-\hat\varepsilon^2,\nn
\\
\Volume\left({\mathcal M}^{2,2}_2(S^1\times I)\right)
 &=&\frac{1}{2}\hat{\mathcal A}^2
 -\left(\frac{5}{3}-\hat\varepsilon+\hat\varepsilon^2\right)\hat{\mathcal A}
 +\frac{17}{12}-\frac{5}{3}\hat\varepsilon+2\hat\varepsilon^2
 -\frac{2}{3}\hat\varepsilon^3+\frac{1}{3}\hat\varepsilon^4,
\nn\\
 \Volume\left({\mathcal M}^{2,2}_3(S^1\times I)\right)
&=&\frac{1}{6}\A^3-\frac{1}{2}\left(\frac{7}{3}-\hat\varepsilon+\hat\varepsilon^2\right)\A^2\nn\\
&&+\left(\frac{331}{120}-\frac{7}{3}\hat\varepsilon+\frac{8}{3}\hat\varepsilon^2-\frac{2}{3}\hat\varepsilon^3
+\frac{1}{3}\hat\varepsilon^4\right)\A\nn\\
&&-\frac{793}{360}+\frac{331}{120}\hat\varepsilon-\frac{85}{24}\hat\varepsilon^2+\frac{29}{18}\hat\varepsilon^3
-\frac{11}{12}\hat\varepsilon^4+\frac{2}{15}\hat\varepsilon^5-\frac{2}{45}\hat\varepsilon^6,
\nn\\
 \Volume\left({\mathcal M}^{2,2}_4(S^1\times I)\right)
&=&\frac{1}{24}\A^4-\frac{1}{6}\left(3-\hat\varepsilon+\hat\varepsilon^2\right)\A^3\nn\\
&&+\frac{1}{2}\left(\frac{409}{90}-3\hat\varepsilon+\frac{10}{3}\hat\varepsilon^2
-\frac{2}{3}\hat\varepsilon^3+\frac{1}{3}\hat\varepsilon^4\right)\A^2\nn\\
&&-\left(\frac{292}{63}-\frac{409}{90}\hat\varepsilon+\frac{111}{20}\hat\varepsilon^2
-\frac{37}{18}\hat\varepsilon^3+\frac{41}{36}\hat\varepsilon^4-\frac{2}{15}\hat\varepsilon^5
+\frac{2}{45}\hat\varepsilon^6\right)\A\nn\\
&&+\frac{18047}{5040}-\frac{292}{63}\hat\varepsilon+\frac{37}{6}\hat\varepsilon^2-\frac{16}{5}\hat\varepsilon^3
+\frac{35}{18}\hat\varepsilon^4-\frac{19}{45}\hat\varepsilon^5+\frac{7}{45}\hat\varepsilon^6-\frac{4}{315}\hat\varepsilon^7+\frac{1}{315}\hat\varepsilon^8,\nn\\
 \label{local vortices}
\eea
at the finite $\hat\varepsilon$.

Taking the limit of $\hat\varepsilon\to0$ in the above 
results of (\ref{local vortices}) for $k=1,2,3,4$, we 
finally obtain the moduli space volume of vortices 
 with $N_f=N_c=2$ on the cylinder $S^1\times I$ 
\bea
\Volume\left({\mathcal M}^{2,2}_0(S^1\times I)\right)
 &=& 1,\nn\\
\Volume\left({\mathcal M}^{2,2}_1(S^1\times I)\right)
 &=&\hat{\mathcal A} -1,\nn\\
\Volume\left({\mathcal M}^{2,2}_2(S^1\times I)\right)
 &=&\frac{1}{2}\hat{\mathcal A}^2
 -\frac{5}{3}\hat{\mathcal A}
 +\frac{17}{12},
\label{local in the e to 0 limit}
\label{eq:local-vortex-cylinder}
\\
 \Volume\left({\mathcal M}^{2,2}_3(S^1\times I)\right)
&=&\frac{1}{6}\A^3-\frac{7}{6}\A^2+\frac{331}{120}\A-\frac{793}{360},
\nn\\
 \Volume\left({\mathcal M}^{2,2}_4(S^1\times I)\right)
&=&\frac{1}{24}\A^4-\frac{1}{2}\A^3+\frac{409}{180}\A^2-\frac{292}{63}\A+\frac{18047}{5040}.
\nn
\eea
Surprisingly, they completely agree with the volume of 
the moduli space of the local vortices on the sphere $S^2$, 
derived in \cite{Miyake:2011yr}, 
up to an overall normalization and 
 a rescaling to define the moduli space. 
The computation of the volume of the moduli space of 
vortices on sphere $S^2$ has given the asymptotic behavior 
at large area $\A$ which reduces drastically when $N_c=N_f$, 
and has suggested a formula \cite{Miyake:2011yr}\footnote{
Eq.(4.52) of Ref.\cite{Miyake:2011yr} has an additional factor 
of $N!$ which we have forgotten to divide out, apart from 
a rescaling by $(2\pi)^N$ to define the moduli space. 
}
\be
\Volume\left({\mathcal M}^{N,N}_k(S^2)\right)\sim
\frac{\A^k}{k!}.
\ee

The physical reason of the reduction of asymptotic power 
of the volume is the following. 
When $N_c=N_f$, the non-Abelian vortices are called 
non-Abelian local vortices, since the field configuration 
approaches the (unique) vacuum exponentially 
\cite{Eto:2009wq} outside of 
the local vortices of the intrinsic size $1/(g^2c)$. 
Their position moduli ($k$ complex dimensions) can extend 
to the entire area, whereas all the other moduli 
($k(N-1)$ complex dimensions) 
correspond to orientations in internal flavor symmetry 
and can spread only up to the size $1/(g^2c)$ around 
the local vortex \cite{Eto:2005yh, Eto:2006pg}. 
Therefore the asymptotic power of $\A$ for local vortices 
is just $k$, corresponding only to the number of the 
position moduli. 
When $N_c < N_f$, on the other hand, vortices are called 
semi-local vortices, since the field configuration approaches 
to (non-unique) vacua only in some powers of the distance 
away from the vortices. 
Not only the position moduli ($k$ complex dimensions) 
but also all the other moduli ($k(N_f-1)$ complex dimensions) 
can now extend to the entire area. 
These $k(N_f-1)$-dimensional moduli are called the size moduli 
instead of the orientational moduli \cite{Eto:2007yv}. 
This is the reason why the asymptotic power of $\A$ 
becomes $kN_f$ for the semi-local vortices. 

From this physical consideration, it is interesting and 
gratifying to see that the volume 
(\ref{eq:local-vortex-cylinder}) of the moduli space of 
the local vortices $N_c=N_f$ on the cylinder 
agrees exactly with that on the sphere $S^2$. 
We also note that the volume on the cylinder 
(\ref{local vortices}) before taking 
the limit $\hat\varepsilon \to 0$ 
can depend on 
the mass difference $\hat\varepsilon$, but only at 
non-leading powers of $\A$. 
Since the mass differences are originated from 
holonomies at the boundaries of the cylinder 
\cite{Eto:2007aw, Eto:2006mz}, 
this result is also consistent with the notion that the 
effect of holonomy only extends up to a finite distance 
from the boundary for local vortices with the intrinsic 
size $1/(g^2c)$. 
So these non-trivial results, including the coefficients of 
the polynomial, suggest that our localization formula and 
T-duality between the domain-walls and vortices 
works correctly.

\begin{figure}
\begin{center}
\begin{tabular}{ccc}
\includegraphics[scale=0.8]{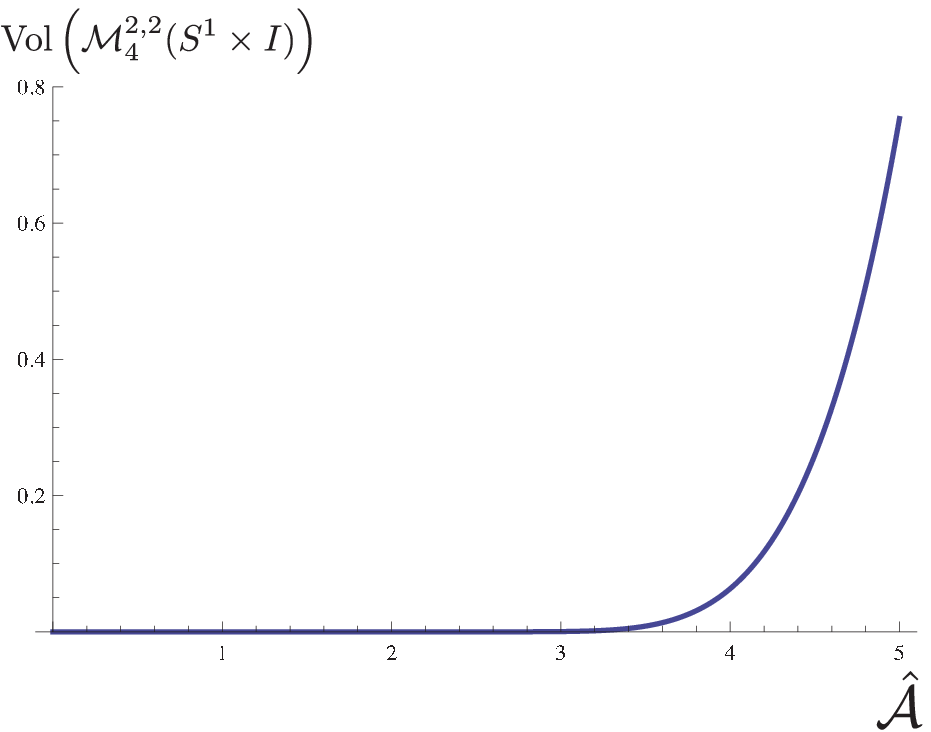}&&
\includegraphics[scale=0.8]{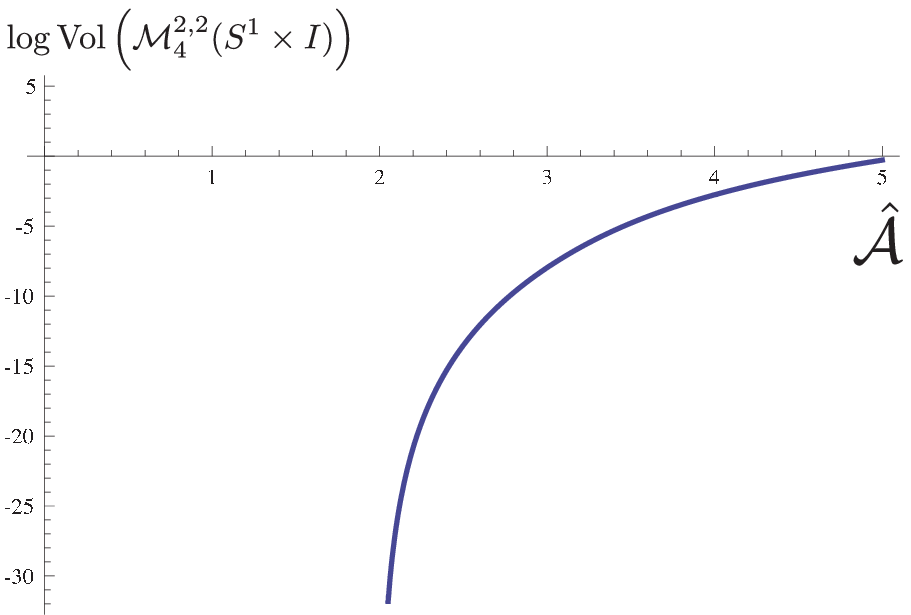}\\
(a) && (b)
\end{tabular}
\end{center}
\caption{
(a) The volume of the moduli space of the non-Abelian local
vortices on the cylinder $S^1\times I$ 
as a function of the area $\A$,
for $N_c=N_f=2$ and $k=4$.
(b) A logarithmic plot of the volume, 
showing that the volume vanishes at $\A=2$ 
(the Bradlow bound).
The volume as a function of $\A$ differs in the regions 
$\A>4$, $3<\A\leq4$, $2<\A\leq3$ and $\A<2$ 
as one approaches the Bradlow bound.
The different functions are smoothly connected with each 
other at the boundaries of each regions 
up to high derivatives.
}
\label{Volume Plot}
\end{figure}

So far, we have assumed that the area $\A$ is sufficiently 
larger than the vortex charge $k$. 
However for the fixed vortex charge $k$, there exists an 
exact lower bound of the area, which is called the Bradlow 
bound \cite{Bradlow:1990ir}. 
The Bradlow bound of the volume essentially comes from the 
integral formula (\ref{contour integral}), 
where the integral vanishes if the exponent is negative.
So the behavior of the volume changes whether the area is 
larger than the charge or not.
As a result, the functional dependence of the volume 
on $\A$ changes as $\A$ decreases towards the Bradlow bound. 
For example, let us consider again the case that 
$N_c=N_f=2$ (local vortex) and $k=4$ 
in the limit of $\hat\varepsilon\to0$.
As explained, if $\A$ is larger than 4, the volume is given 
in (\ref{local in the e to 0 limit}). 
If $3< \A \leq4$, then all the terms containing the factor 
$(\A-4)$ (in the limit of $\hat\varepsilon\to 0$) 
in (\ref{k=4 A>4}) drop out 
because of the formula (\ref{contour integral}).
Then the volume becomes
\be
 \Volume\left({\mathcal M}^{2,2}_4(S^1\times I)\right)=
\frac{\A^8}{6720}-\frac{\A^7}{252}+\frac{2\A^6}{45}-\frac{4\A^5}{15}+\frac{67\A^4}{72}-\frac{173\A^3}{90}
+\frac{409\A^2}{180}-\frac{436\A}{315}+\frac{1663}{5040}.
\ee
If $2 < \A \leq 3$, we find similarly 
\be
 \Volume\left({\mathcal M}^{2,2}_4(S^1\times I)\right)=
\frac{\A^8}{2880}-\frac{\A^7}{180}+\frac{7\A^6}{180}-\frac{7\A^5}{45}+\frac{7\A^4}{18}-\frac{28\A^3}{45}
+\frac{28\A^2}{45}-\frac{16\A}{45}+\frac{4}{45}.
\ee 
The volume vanishes if $\A\leq 2$. 
We plot the volume as a function of $\A$ for the above 
regions in Fig.~\ref{Volume Plot}.
We note that the functions are smoothly connected at each 
boundary ($\A=3$ and $4$), since the derivatives coincide 
with each other up to high orders.

\section{Conclusion and Discussion}

In this paper, we have formulated a path-integral to 
obtain the volume of the moduli space of the domain-walls. 
We have seen that the localization method is a powerful 
tool to calculate the volume of the moduli space 
without the explicit metric. 
We have also noticed that the localization method is 
useful to understand not only the global structure of 
the moduli space like the volume, but also the detailed 
and interesting properties of the moduli space 
through the dualities. 

So far, we have not assumed that the supersymmetry is 
behind the BPS domain-wall system. 
However, the BRST symmetry, which plays important roles 
in the localization method, is known to be regarded as 
a part of the supersymmetry. 
Actually our BRST transformations (\ref{BRSTvec}) and 
(\ref{BRSTchi}) are the dimensional reduction 
of the two-dimensional A-twisted supersymmetric 
transformation to one dimension. 
So we can expect that our volume formula is closely 
related to a partition function or vacuum expectation 
value (vev) in supersymmetric gauge theories. 
It is interesting to explore non-perturbative 
corrections in supersymmetric gauge theories 
from the viewpoint of the volume formula of the 
moduli space of the BPS domain-walls. 
When boundaries are present, in particular, 
not much is known for the non-perturbative corrections in 
supersymmetric gauge theories. 
We have found that the boundary conditions are important 
and determine various interesting properties of 
the volume calculation. 
The volume calculation of the moduli space in 
supersymmetric gauge theories with the boundaries may 
shed light on the non-perturbative dynamics and dualities.

We have obtained the exact results of the volume of 
the moduli space by using the localization method,
but more directly we can also obtain the volume from 
an integral of a volume form, constructed by the 
explicit metric, over the moduli space.
The volume is an integral result, where the local 
information is smeared out, but we can expect that 
informations on the local metric can be reconstructed 
from the various uses of the localization method.

We sometimes encounter a mysterious relationship between 
the BPS solitons and (quantum mechanical) integrable 
systems like spin chains.
The partition functions and vevs in supersymmetric gauge 
theories often become 
important quantities in the integrable systems. 
Our integral formula for the volume of the moduli space, 
which is expressed in terms of the determinant of 
the transition matrix, is also reminiscent of the 
integrable systems. 
We would like to investigate the relationship between 
the volume calculation of the BPS 
solitons and integrable systems in the future.

The volume of the moduli space is also mathematically 
interesting since the localization method says that the 
volume is almost determined by a topological nature 
of the moduli space. 
The volume of the moduli space may express a topological 
invariants of the moduli spaces. 
Recently the localization of the $\mathcal{N}=(2,2)$ 
supersymmtric gauge theories on $S^2$ have been 
performed \cite{Benini:2012ui,Doroud:2012xw}.
The partition function of the $\mathcal{N}=(2,2)$ 
supersymmtric gauge theories has two alternative 
expressions. 
One uses the localization around the Higgs branch, 
where the partition function reduces to the (anti-)vortex 
moduli zero-modes theory known as the (anti-)vortex 
partition function 
\cite{Shadchin:2006yz,Dimofte:2010tz,Yoshida:2011au,
Miyake:2011yr, Bonelli:2011fq,Fujimori:2012ab}.
The other uses the localization around the Coulomb 
branch, where the path integral reduces to the 
multi-contour integrals. 
These two expressions turn out to be identical. 
Moreover, it is conjectured in \cite{Jockers:2012dk} 
(see also \cite{Gomis:2012wy}) that  the free energy 
of the $\mathcal{N}=(2,2)$ supersymmtric gauge theories 
is  the quantum (world sheet instanton) corrected 
\Kahler potential of  \Kahler moduli for the Higgs 
branch and actually reproduces the genus-zero 
Gromov-Witten invariant which counts holomorphic 
maps from the sphere to the target space manifold.

We have investigated the volume of the moduli space of 
the vortices on the cylinder via the T-duality. 
So we can expect that our vortex results on the 
cylinder may produce the moduli space of novel 
holomorphic maps from the cylinder to the target 
manifold.

The width of domain-walls in an infinite interval 
has been studied in detail. 
If the mass difference of scalar fields $H$ 
taking non-vanishing values in the two adjacent vacua 
is denoted as $\Delta m$, the width of the domain-wall 
is given by $|\Delta m|/(g^2c)$ in the weak coupling 
region ($g\sqrt{c} \ll |\Delta m|$), but by 
$1/g\sqrt{c}$ in the strong coupling region 
($g\sqrt{c} \gg |\Delta m|$)
\cite{Kaplunovsky:1998vt, Eto:2006ng, Shifman:2002jm}. 
Our results from the localization formula are consistent 
with the weak coupling result for the infinite interval. 
Therefore our results suggest that the width of the 
domain-wall for finite interval does not change 
significantly as we move from weak coupling toward 
strong coupling region. 
Since the effect of boundary is stronger as the length 
of interval decreases, it is quite possible that 
the intuition gained from the infinite interval 
case is not valid for domain-walls in short intervals. 
It is an interesting future problem to work out 
the domain-wall solution at finite (short) interval 
carefully.

We had to guess the sign factors associated with the 
intersection number of color-lines. 
We can guess that the sign factor may originate from 
the fact that our diagonal gauge fixing condition 
$\Phi^\alpha=0$ is ambiguous and ill-defined when 
eigenvalues $\phi_a$ of the matrix $\Phi$ are 
degenerate. 
The color-line connecting the boundary conditions 
at left and right boundaries are usually formulated 
in terms of eigenvalues of the matrix $\Sigma$, 
which is canonically conjugate to $\Phi$. 
This complication is one of the reasons that prevented 
us to derive more explicitly the sign factors 
from the precise treatment of path-integral. 
We leave this question for a future study.

\section*{Acknowledgment}

This work is supported in part by Grant-in-Aid for 
Scientific Research from the Ministry of Education, Culture, 
Sports, Science and Technology, Japan No.21540279 (N.S.), 
No.21244036 (N.S.).

\appendix

\section{Explicit computation of $N_c=1$ and $N_c=2$ 
with $N_f=3$}\label{app:explicit-comp}

Using Eqs.~(\ref{dual1:Abelian volume}) and 
(\ref{dual1:non-Abelian volume}) for $N_f=3$, 
we find that our localization formula gives 
\be
\Volume\left({\mathcal M}^{1,3}_{1\to 3}\right)
=\frac{\beta^{2}}{2}
(\hat{L}-d_{13})^{2},
\label{eq:Abelian volume1}
\ee
\be
\Volume\left({\mathcal M}^{2,3}_{(1,2) 
\to (2,3)}\right)=
\frac{\beta^{2}}{2}(\hat{L}^2-d_{13}^2
+2d_{12}d_{23}).
\label{eq:non-Abelian volume1}
\ee
They differ already at the next-to-leading order in 
$\hat{L}$. 

To check our results of localization formula 
at non-leading powers of $\hat{L}$, let us compute the volume 
using the rigid-rod approximation 
\cite{Eto:2007aw} where the domain-wall 
connecting masses $m_i$ and $m_j$ has a width $d_{ij}$. 
Let us denote the position of 
the first (second) wall as $y_1$ ($y_2$). 
For the Abelian gauge theory $N_c=1$, 
two walls are non-penetrable \cite{Isozumi:2004jc,Eto:2006ng}. Therefore we obtain 
\bea
\Volume\left({\mathcal M}^{1,3}_{1\to 3}\right)
&=&\beta^{2}
\int_{\frac{d_{12}}{2}}^{\hat{L}-d_{23}-\frac{d_{12}}{2}}
dy_1 
\int_{y_1+\frac{d_{13}}{2}}^{\hat{L}-\frac{d_{23}}{2}}
dy_2\nn\\
&=&\frac{\beta^{2}}{2}(\hat{L}-d_{13})^2,
\label{eq:Abelian volume2}
\eea
giving an identical result as our localization formula 
(\ref{eq:Abelian volume1}). 
For non-Abelian gauge theory $N_c=2$, two domain walls 
are also non-penetrable, but the allowed region of positions 
are different. We separate the integration region into 
two and obtain 
\bea
\Volume\left({\mathcal M}^{2,3}_{(1,2) 
\to (2,3)}\right)&=&
\beta^{2}
\left(
\int_{\frac{d_{23}}{2}+d_{12}}^{\hat{L}-\frac{d_{23}}{2}}
dy_1 
\int_{y_1-\frac{d_{13}}{2}}^{\hat{L}-\frac{d_{12}}{2}}
dy_2 
+
\int_{\frac{d_{23}}{2}}^{\frac{d_{23}}{2}+d_{12}}
dy_1 
\int_{\frac{d_{12}}{2}}^{\hat{L}-\frac{d_{12}}{2}}
dy_2 \right)\nn\\
&=&\frac{\beta^2}{2}(\hat{L}-2d_{12}+d_{13})(\hat{L}-d_{13})
+\beta^2(L-d_{12})d_{12}\nn\\
&=&\frac{\beta^{2}}{2}(\hat{L}^2-d_{13}^2
+2d_{12}d_{23}),
\label{eq:non-Abelian volume2}
\eea
giving an identical result as our localization formula 
(\ref{eq:non-Abelian volume1}).

\section{Volume of Moduli Space of Dual Non-Abelian Domain 
Walls \label{app:nonabelian-duality}
} 

We consider the topological sector with the maximal 
number of domain walls in $U(N_c)$ gauge theories with 
$N_f$ flavors of scalar fields in the fundamental 
representation. 
The volume of the moduli space of 
domain walls is given by the determinant of the 
transition matrix ${\mathcal T}^{N_c,N_f}_{(1,\cdots,N_c) 
\to (\tilde N_c,\cdots,N_f)}$ as 
\begin{equation}
\Volume\left({\mathcal M}^{N_c,N_f}_{(1,\dots,N_c) 
\to (\tilde N_c,\cdots,N_f)}\right)
=\beta^{N_f} \Det {\mathcal T}^{N_c,N_f}_{(1,\cdots,N_c) 
\to (\tilde N_c,\cdots,N_f)}. 
\end{equation}
The leading behavior at large volume is given 
by the largest powers in $\hat L$ as 
\be
\lim_{\hat{L}\to\infty}\frac{{\mathcal T}^{N_c,N_f}_{(1,\cdots,N_c) \to 
(\tilde N_c,\cdots,N_f)}}{\hat{L}^{N_f}}
=
\begin{pmatrix}
\frac{1}{\tilde{N}_c!} & \cdots & \frac{1}{(N_c-2)!} 
& \frac{1}{(N_c-1)!} \\
\vdots & \ddots & \vdots & \vdots \\
\frac{1}{(\tilde N_c-N_c+2)!}  & \cdots & \frac{1}{\tilde{N}_c!} 
& \frac{1}{(\tilde N_c+1)!} \\
\frac{1}{(\tilde N_c-N_c+1)!} & \cdots & \frac{1}{(\tilde N_c-1)!} 
& \frac{1}{\tilde N_c!}
\end{pmatrix}.
\ee
Let us define the determinant of the matrix in the right-hand side 
as $\Delta^{N_c, N_f}$. 
For $N_c > \tilde N_c$, the above formula contains factorials of 
negative integer at the lower left corner. 
These factorials should be interpreted as zeros 
\be
\frac{1}{(-n)!}=\frac{1}{\Gamma(-n+1)}=0, \qquad 
n\in \mathbb{Z}_+. 
\ee
In order to obtain the determinant, we subtract the 
$(N_c-1)$-th row multiplied by $\tilde N_c+1$ 
from the $N_c$-th (last) row of the right-hand side 
in order to eliminate the right-most entry of the $N_c$-th row 
\be
\Delta^{N_c, N_f}=
\Det
\begin{pmatrix}
\frac{1}{\tilde{N}_c!} & \cdots & \frac{1}{(N_c-2)!} 
& \frac{1}{(N_c-1)!} \\
\vdots & \ddots & \vdots & \vdots \\
\frac{1}{(\tilde N_c-N_c+2)!}  & \cdots & \frac{1}{\tilde{N}_c!} 
& \frac{1}{(\tilde N_c+1)!} \\
\frac{-(N_c-1)}{(\tilde N_c-N_c+1)!} & \cdots 
& \frac{-1}{(\tilde N_c-1)!} 
& 0
\end{pmatrix}.
\ee
Similarly subtracting the $(N_c-2)$-th row multiplied by 
$\tilde N_c+2$ from the $(N_c-1)$-th row and continuing 
the procedure, we can eliminate all the entries of the 
$N_c$-th column except the first row. Thus we find 
\be
\Delta^{N_c, N_f}=
\Det
\begin{pmatrix}
\frac{1}{\tilde{N}_c!} & \cdots & \frac{1}{(N_c-2)!} 
& \frac{1}{(N_c-1)!} \\
\frac{-(N_c-1)}{(\tilde N_c-1)!}  & \cdots 
& \frac{-1}{(N_f-2)!} & 0 \\
\vdots & \ddots & \vdots & \vdots \\
\frac{-(N_c-1)}{(\tilde N_c-N_c+1)!} & \cdots 
& \frac{-1}{(\tilde N_c-1)!}
& 0
\end{pmatrix}.
\ee
Therefore we obtain the recursion relation 
\be
\Delta^{N_c, N_f}=
\frac{(N_c-1)!}{(N_f-1)!}\Delta^{N_c-1, N_f-1}.
\ee
The recursion relation is solved with the initial 
condition $\Delta^{1, N_f}=1/(N_f-1)!$ to give 
\be
\Delta^{N_c, N_f}
=\frac{\prod_{j=1}^{N_c}(j-1)! \times 
\prod_{k=1}^{\tilde{N}_c}(k-1)!}
{\prod_{i=1}^{N_f}(i-1)!}.
\ee
Thus we find the duality (\ref{eq:leading_duality}) 
is valid. 
Moreover the coefficient of the leading term 
is given by the volume of the Grassmann manifold 
(\ref{eq:grassmann_volume}) 
apart from the intrinsically ambiguous overall normalization 
factor to define the moduli space. 
The proof here is valid also for the leading behavior of 
the equivalence of Abelian and non-Abelian domain walls, 
namely agreement between Eqs.~(\ref{vol of CP 1}) and 
(\ref{vol of CP 2}).


\begin{thebibliography}{99}

\bibitem{Manton:1993tt}
  N.~S.~Manton,
  Nucl.\ Phys.\  B {\bf 400} (1993) 624.

\bibitem{Shah:1993us}
  P.~A.~Shah and N.~S.~Manton,
  J.\ Math.\ Phys.\  {\bf 35} (1994) 1171
  [arXiv:hep-th/9307165].

\bibitem{Manton:1998kq}
  N.~S.~Manton and S.~M.~Nasir,
  Commun.\ Math.\ Phys.\  {\bf 199} (1999) 591
  [arXiv:hep-th/9807017].

\bibitem{Nasir:1998kt}
  S.~M.~Nasir,
  Phys.\ Lett.\  B {\bf 419} (1998) 253
  [arXiv:hep-th/9807020].

\bibitem{Manton:2004tk}
  N.~S.~Manton and P.~Sutcliffe,
  ``Topological solitons,''
{\it  Cambridge, UK: Univ. Pr. (2004) 493 p}.

\bibitem{Nekrasov:2002qd}
  N.~A.~Nekrasov,
  Adv.\ Theor.\ Math.\ Phys.\  {\bf 7} (2004) 831
  [arXiv:hep-th/0206161].

\bibitem{Moore:1997dj}
  G.~W.~Moore, N.~Nekrasov and S.~Shatashvili,
  Commun.\ Math.\ Phys.\  {\bf 209} (2000) 97
  [arXiv:hep-th/9712241].

\bibitem{Gerasimov:2006zt}
  A.~A.~Gerasimov and S.~L.~Shatashvili,
  Commun.\ Math.\ Phys.\  {\bf 277} (2008) 323
  [arXiv:hep-th/0609024].

\bibitem{Miyake:2011yr}
  A.~Miyake, K.~Ohta and N.~Sakai,
  Prog.\ Theor.\ Phys.\  {\bf 126} (2012) 637
  [arXiv:1105.2087 [hep-th]];
  J.\ Phys.\ Conf.\ Ser.\  {\bf 343} (2012) 012107
  [arXiv:1111.4333 [hep-th]].

\bibitem{Isozumi:2004va}
  Y.~Isozumi, M.~Nitta, K.~Ohashi and N.~Sakai,
  Phys.\ Rev.\ D {\bf 70} (2004) 125014
  [hep-th/0405194].

\bibitem{Antoniadis:1996ra}
  I.~Antoniadis and B.~Pioline,
  Int.\ J.\ Mod.\ Phys.\ A {\bf 12} (1997) 4907
  [hep-th/9607058].

\bibitem{Eto:2004vy}
  M.~Eto, Y.~Isozumi, M.~Nitta, K.~Ohashi, K.~Ohta and N.~Sakai,
  Phys.\ Rev.\ D {\bf 71} (2005) 125006
  [hep-th/0412024].

\bibitem{Eto:2007aw}
  M.~Eto, T.~Fujimori, M.~Nitta, K.~Ohashi, K.~Ohta and N.~Sakai,
  Nucl.\ Phys.\ B {\bf 788} (2008) 120
  [hep-th/0703197].



\bibitem{Isozumi:2004jc}
  Y.~Isozumi, M.~Nitta, K.~Ohashi and N.~Sakai,
  Phys.\ Rev.\ Lett.\  {\bf 93} (2004) 161601
  [hep-th/0404198].

\bibitem{Eto:2006ng}
  M.~Eto, Y.~Isozumi, M.~Nitta, K.~Ohashi and N.~Sakai,
  hep-th/0607225.

\bibitem{Shifman:2002jm}
  M.~Shifman and A.~Yung,
  Phys.\ Rev.\ D {\bf 67} (2003) 125007
  [hep-th/0212293].


\bibitem{Callias:1977kg} 
  C.~Callias,
  Commun.\ Math.\ Phys.\  {\bf 62}, 213 (1978).

\bibitem{AtiyahPatodiSinger}
M.~F.~Atiyah, V.~K.~Patodi, and I.~M.~Singer,  
Math.~Proc.~Camb.~Phil.~Soc.~{\bf 77}, 43 (1975). 

\bibitem{Macdonald}
I. G. Macdonald, 
Invent. Math. {\bf 56} (1980), 93.

\bibitem{Fujii}
K.~Fujii,
J. Appl. Math. {\bf 2} (2002), 371-405.

\bibitem{Ooguri:2002gx}
  H.~Ooguri and C.~Vafa,
  Nucl.\ Phys.\  B {\bf 641} (2002) 3
  [arXiv:hep-th/0205297].

\bibitem{Eto:2006mz}
  M.~Eto, T.~Fujimori, Y.~Isozumi, M.~Nitta, K.~Ohashi, K.~Ohta and N.~Sakai,
  Phys.\ Rev.\ D {\bf 73} (2006) 085008
  [arXiv:hep-th/0601181].

\bibitem{Abrikosov:1956sx}
A.~A.~Abrikosov,
``On The Magnetic Properties Of Superconductors Of The Second Group,''
Sov.\ Phys.\ JETP {\bf 5} (1957) 1174 
[Zh.\ Eksp.\ Teor.\ Fiz.\  {\bf 32} (1957) 1442]; 
H.~B.~Nielsen and P.~Olesen,
``Vortex-Line Models For Dual Strings,''
Nucl.\ Phys.\ {\bf B61} (1973) 45.

\bibitem{Eto:2008dm}
  M.~Eto, T.~Fujimori, M.~Nitta, K.~Ohashi and N.~Sakai,
  Phys.\ Rev.\ D {\bf 77} (2008) 125008
  [arXiv:0802.3135 [hep-th]].

\bibitem{Eto:2009wq}
  M.~Eto, T.~Fujimori, T.~Nagashima, M.~Nitta, K.~Ohashi and N.~Sakai,
  Phys.\ Lett.\ B {\bf 678} (2009) 254
  [arXiv:0903.1518 [hep-th]].

\bibitem{Eto:2005yh}
  M.~Eto, Y.~Isozumi, M.~Nitta, K.~Ohashi and N.~Sakai,
  Phys.\ Rev.\ Lett.\  {\bf 96} (2006) 161601
  [hep-th/0511088].

\bibitem{Eto:2006pg}
  M.~Eto, Y.~Isozumi, M.~Nitta, K.~Ohashi and N.~Sakai,
  J.\ Phys.\ A {\bf 39} (2006) R315
  [hep-th/0602170].

\bibitem{Eto:2007yv}
  M.~Eto, J.~Evslin, K.~Konishi, G.~Marmorini, M.~Nitta, K.~Ohashi, W.~Vinci and N.~Yokoi,
  Phys.\ Rev.\ D {\bf 76} (2007) 105002
  [arXiv:0704.2218 [hep-th]].

\bibitem{Bradlow:1990ir}
  S.~B.~Bradlow,
  Commun.\ Math.\ Phys.\  {\bf 135} (1990) 1.

\bibitem{Benini:2012ui} 
  F.~Benini and S.~Cremonesi,
  arXiv:1206.2356 [hep-th].
  
\bibitem{Doroud:2012xw} 
  N.~Doroud, J.~Gomis, B.~Le Floch and S.~Lee,
  arXiv:1206.2606 [hep-th].

\bibitem{Shadchin:2006yz} 
  S.~Shadchin,
  JHEP {\bf 0708}, 052 (2007)

\bibitem{Dimofte:2010tz} 
  T.~Dimofte, S.~Gukov and L.~Hollands,
  Lett.\ Math.\ Phys.\  {\bf 98}, 225 (2011)

\bibitem{Yoshida:2011au} 
  Y.~Yoshida,
  arXiv:1101.0872 [hep-th].

\bibitem{Bonelli:2011fq} 
  G.~Bonelli, A.~Tanzini and J.~Zhao,
  JHEP {\bf 1206}, 178 (2012)

\bibitem{Fujimori:2012ab} 
  T.~Fujimori, T.~Kimura, M.~Nitta and K.~Ohashi,
  JHEP {\bf 1206}, 028 (2012)

\bibitem{Jockers:2012dk} 
  H.~Jockers, V.~Kumar, J.~M.~Lapan, D.~R.~Morrison and M.~Romo,
  arXiv:1208.6244 [hep-th].

\bibitem{Gomis:2012wy} 
  J.~Gomis and S.~Lee,
  arXiv:1210.6022 [hep-th].

\bibitem{Kaplunovsky:1998vt}
  V.~S.~Kaplunovsky, J.~Sonnenschein and S.~Yankielowicz,
  Nucl.\ Phys.\ B {\bf 552} (1999) 209
  [hep-th/9811195].

 \end{thebibliography}
\end{document}